\documentclass[]{aa}
\usepackage{graphicx}
\usepackage{txfonts}
\usepackage{xcolor}
\usepackage{hyperref}
\hypersetup{
  colorlinks,
  citecolor=blue,
  linkcolor=red,
  urlcolor=blue}

\begin{document}

   \title{A systematic method to identify runaways from star clusters produced from single-binary interactions:}

   \subtitle{A case study of M67}

    \author{A. Herrera-Urquieta
          \inst{1}
          , N. Leigh
          \inst{1,2}
          , J. Pinto
          \inst{1}
          , G. Díaz-Cerda
          \inst{3}
          , S. M. Grondin
          \inst{4}\\
          J. J. Webb 
          \inst{4}
          , R. Mathieu 
          \inst{5}
          , T. Ryu
          \inst{6,7}
          , A. Geller 
          \inst{8}
          , M. Kounkel 
          \inst{9}
          , S. Toonen
          \inst{10}
          \and
          M. Vilaxa-Campos
          \inst{1}
          }

   \institute{Departamento de Astronom\'ia, Facultad Ciencias F\'isicas y Matem\'aticas, Universidad de Concepci\'on, Chile
        \and
             Department of Astrophysics, American Museum of Natural History, Central Park West and 79th Street, New York, NY, 10024, USA
        \and 
             Instituto Profesional Dr. Virginio Gómez, Arturo Prat 196, 4070298 Concepción, Bío Bío
        \and
             David A. Dunlap Department of Astronomy and Astrophysics, University of Toronto, 50 St. George Street, Toronto, ON, M5S 3H4, Canada
        \and
            Department of Astronomy, University of Wisconsin-Madison, 475 North Charter Street, Madison, WI 53706, USA
        \and 
             Max-Planck-Institut für Astrophysik, Karl-Schwarzschild-Straße 1, 85748 Garching bei München, Germany
        \and
            Physics and Astronomy Department, Johns Hopkins University, Baltimore, MD 21218, USA
        \and
            Center for Interdisciplinary Exploration and Research in Astrophysics (CIERA) and Department of Physics and Astronomy Northwestern University, 1800 Sherman Ave, Evanston, IL 60201 USA
        \and
            Department of Physics and Astronomy, Vanderbilt University, Nashville, TN 37235, USA
        \and
            Anton Pannekoek Institute for Astronomy, University of Amsterdam, Science Park 904, 1098 XH Amsterdam, Netherlands}

   \authorrunning{A.\ Herrera-Urquieta et al.}
   \date{Received MM DD, YY; accepted MM DD, YY}

 
  \abstract
   {One hypothesis for runaway stars (RSs) is that they are ejected from star clusters with high velocities relative to the cluster center-of-mass motion. There are two competing mechanisms for their production: supernova-based ejections in binaries, where one companion explodes, leaves no remnant, and launches the other companion at the instantaneous orbital velocity, and the disintegration of triples (or higher-order multiples), which produces a recoiled runaway binary (RB) and an RS.}
   {We search for RS candidates using data from the Gaia DR3 survey with a focus on triple disintegration since in this case the product is always a binary and a single star that should be moving in opposite directions.}
   {We created a systematic methodology to look for candidate RS-RB runaway pairs produced from the disintegration of bound three-body systems formed from single-binary interactions based on momentum conservation and causality. The method we use is general and can be applied to any cluster with a 5D kinematic data set. We used our criteria to search for these pairs in a 150 pc circular field of view surrounding the open cluster M67, which we used as a benchmark cluster to test the robustness of our method.}
   {Our results reveal only one RS-RB pair that is consistent with all of our selection criteria out of an initial sample of $10^8$ pairs.}
   {}


   \titlerunning{A systematic method to identify runaways}
   \authorrunning{A. Herrera Urquieta et al.}
   \maketitle
%

\section{Introduction}

Star clusters with large stellar densities $\gtrsim 100$ M$_{\odot}$pc$^{-3}$ create very active dynamical environments where binary and single stars interact on short timescales of order a few megayears \citep[e.g.,][]{Leigh_2011, Leigh_2012}.  Such star clusters evolve due to intracluster dynamics \citep{Reipurth_2010} and eject stars into the Galactic field via a combination of two-body relaxation and the disintegration of higher-order multiples (usually binaries) interacting (often chaotically) due to single-binary and binary-binary interactions. This binding and subsequent dissociation of temporary groupings of three or more stars occurs in all star clusters, including nuclear star clusters, globular clusters (GCs), and open clusters (OCs).  Analogously, it is also possible for stable triple systems to become unstable and eject a binary star and a single star.  We expect these events to be relatively rare. Specifically, only 2-4\% of all low- and intermediate-mass triples become unstable, and of those, 42-45\% lead to an unbound escaper, while the rest lead to a merger event \citep{Toonen_2022}.
For especially young clusters, stars can also be ejected due to supernova (SN) explosions in binaries \citep{Perets_2012}, which act to destroy the SN progenitor and launch its binary companion at the instantaneous orbital velocity.

Fast stars ejected from a cluster due to either of the above mechanisms (i.e., excluding two-body relaxation and hence tidal stripping) are known as runaway stars (RSs). For the purposes of this paper, we call any object ejected from a cluster due to either the disintegration of triples (formed primarily from single-binary interactions) or the SN-induced ejection scenario an RS object. Typical velocities of RSs escaping from OCs must be sufficiently high to identify them as RSs, but also low enough that they remain sufficiently close to their host cluster to have a non-negligible probability of being associated with it. Observationally, RSs are typically defined as stars with velocities above 30 km s $^{-1}$ \citep[e.g.,][]{Blauw_1961,Renzo2019} relative to the host cluster center-of-mass velocity, which is a factor of a few greater than the typical GC escape velocity.  But they can have lower velocities as well (i.e., corresponding to ejection velocities roughly equivalent to the escape velocity), in which case they are typically referred to as walkaway stars \citep{Schoettler_2020}.  

Fast-moving stars have been the subject of extensive studies using observations and simulations. There have been many studies on dynamically ejected massive stars from young star clusters using numerical integrations (\citealp{Perets_2012}, \citealp{Fujji_2014}, \citealp{Oh_2015}, \citealp{Andersson_2020}, \citealp{dallamico2021}). In most cases, the studies were performed with few-body scattering experiments (e.g., \citealp{Oh_2015}, \citealp{kroupa2006}, \citealp{gualandris2011}, \citealp{Gualandris2014},\citealp{Taeho_2017_1},\citealp{Taeho_2017_2},\citealp{Taeho_2017_3}) and focused on young star clusters. Interestingly, however, the contribution of RSs coming from old star clusters relative to the total population of observed fast-moving stars in our Galaxy has been argued to be non-negligible, mostly due to the ejection of singles during three-body interactions (e.g., \citealp{weatherford23}, \citealp{Steffani_2023_a}, \citealp{Steffani_2023_b}). Observationally, studies of RSs were initially limited to bright massive stars. However, this changed with the launch of the Gaia mission \citep{gaiadr3_2023}, which has provided precise measurements of positions and velocities for more than 1.5 billion stars in the Milky Way. As a result, numerous new RSs have been discovered (e.g., \citealp{Carretero_Castrillo_2023}, \citealp{Liao_2023}, \citealp{Li_2023}, \citealp{Igoshev_2023}), including ones at the low-mass end, highlighting the fact that they exist across all stellar masses.

The first scenario for producing RSs involves an SN in a binary system \citep{Blauw_1961}. In a single degenerate system (i.e., a compact object plus a non-degenerate companion), the primary star will explode, with the consequent mass loss causing the companion to be launched at the instantaneous orbital velocity \citep[e.g.,][]{Tauris_2014,Renzo2019,Leigh2020}. In a double degenerate system (i.e., those containing two white dwarfs), a dynamically driven double-degenerate double-detonation (D$^{6}$) scenario could occur \citep{2018ApJ...854...52S, 2018ApJ...865...15S}. Here, unstable mass transfer leads to the coalescence of the binary, yielding a Type Ia SN. However in the D$^{6}$ scenario, instead of being fully destroyed during the SN, the companion WD survives and becomes a hypervelocity star ejected at a velocity greater than $1000$ km s$^{-1}$. Recently, there have been large efforts to identify hypervelocity stars produced from the D$^{6}$ scenario \citep{2018ApJ...865...15S, 2023OJAp....6E..28E}. Associating these stars with stellar clusters is challenging, however, as the extreme speeds would cause these stars to travel far from their birth clusters.

The second mechanism for RS production thought to operate in old star clusters is the decay of gravitationally bound systems of three or four stars. There are two ways to create an RS star via a three-(or four-) body disintegration. In the first scenario, a hierarchical triple (or quadruple) is born stable but later becomes unstable due to stellar or binary evolution or due to even internal secular dynamical processes (e.g., Lidov-Kozai cycles) \citep[e.g.,][]{Leigh2020, Toonen_2021, Hamers_2021}.  In the second scenario, three (or four) bodies meet and become gravitationally bound temporarily via a single-binary (or binary-binary) interaction. In both scenarios, the triple disintegrates, launching one RS and leaving behind a runaway binary (RB) with a recoil velocity decided by linear momentum conservation \citep{valtonen_karttunen_2006,Leigh_2018}. Hence, given an observed RS and using the available observational data, it should be possible to compute exactly where the RB is if it formed from the disintegration of a triple \citep[e.g.,][]{Reipurth_2010}. It is then possible to search for the RB in large stellar catalogs. Throughout this work, we focus on triple disintegrations formed via single-binary interactions, which dominate over other types of interactions in dense clusters \citep{Leigh_2011}.

In this paper, we propose a novel observational methodology for the identification of RS-RB pairs formed from the decay of three-body systems due to single-binary interactions in a star cluster.
As we show, it is these RSs that should be identifiable as being associated with a progenitor cluster with the highest confidence relative to other methods that are only able to work with a single RS (e.g., the binary SN-based RS formation mechanism).

We leverage the extensive Gaia database to identify the most probable RS-RB pairs.  We focus in this paper on the old OC M67 (NGC 2682) as a case study, given its close distance to the Sun, moderate age, and dynamically active environment. Located at about 889 pc from the Sun \citep{Cantat_2020}, the old OC M67  is made up of stars of solar age and composition with an isochronal age of $\sim$ 4.0 Gyr \citep{Vanderberg_2004, Balaguer_2007, Viani_2017}.  Proper-motion and radial-velocity surveys of M67 \citep{Yadav_2008, Pasquini_2011, Geller_2015} have confirmed kinematic cluster memberships and estimated the population size to be on the order of 1000. Studies have shown that close stellar encounters involving binary stars should be frequent (i.e., $\gtrsim$ 1 Myr$^{-1}$; see the next section for a more detailed calculation) in M67 \citep{Leonard_1996, Hurley_2005, Leigh_2011} such that we expect to be able to find RS-RB pairs in the field of view surrounding the cluster \citep[e.g.,][]{leigh_2016b}.

A method for finding RS-RB pairs and identifying their progenitor cluster has several important applications. First, in principle, detected RS-RB pairs can be used to say something about what must be going on dynamically inside the cluster. For example, given enough RS-RB pairs, particularly when applying the method to additional clusters, if the current orbital separation of the RBs can be measured, then energy conservation can be applied to constrain the initial binary orbital separation. In turn, it could be possible to place constraints on the underlying distribution of binary orbital separations, for example, by addressing whether or not the observed RS-RB properties are anomalous or expected for a given assumed distribution of binary orbital separations.

Runaway binaries also carry essential information about binary evolution. Three-body interactions can harden or soften the binary orbits; therefore, the properties of an ejected binary (i.e., orbital period, separation, and eccentricity) may be different from its state when it was in the cluster \citep{Leigh_2024}. In the case of compact-object binaries, interactions that lead to compact separations may provide information about the formation channels of transient phenomena, such as Type Ia SN in the case of a white dwarf, or gravitational wave progenitors involving neutron stars or BHs. While many of those events are commonly associated with binary evolution pathways, such as the common envelope channel for compact binary formation (\citealt{Paczynski_1976, Grondin_2024}), the identification of dynamically formed short-period binaries would yield complementary evidence that could be used to disentangle which dynamical or evolutionary processes provide the main contribution to such compact systems. These results point to a broader perspective of the RBs as key tools for probing cluster dynamics, binary evolution, and the interaction between them. That is, we can potentially identify events commonly associated with pure binary evolution but that may have some clear contribution from dynamics, thereby facilitating our understanding of how typical exotica may have formed and to what extent their origins are owed to dynamics.

We also wish to test our theoretical understanding of three-body disintegration and hence the chaotic nature of the three-body problem. That is, we theoretically know the outcome distributions (see \citet{valtonen_karttunen_2006}, \citet{StoneLeigh19}, etc.), but we have never been able to test if these are consistent with data. To determine this, we can ideally generate a larger sample size by looking at other clusters. Such a study becomes most interesting if we are looking for specific outcomes, such as those producing X-ray binaries. The rates will depend sensitively on the number of compact objects in a cluster, how many of these objects are in binaries, and so on. Therefore, the number of expected pairs depends on quantities that are not always directly observable but can be constrained using observations of runaway single-binary pairs. Hence, we obtain an independent measure of the underlying frequency and properties of the compact multiple star systems in a cluster relative to, for example, radial velocity surveys.

Following this initial study, a potential future application lies in the detailed study of black hole (BH)-main sequence (MS) binaries ejected from clusters. We can investigate the frequency of BHs in GCs since the predicted ejection rates depend on how many BHs are in a given cluster.  Hence, this offers an independent means of constraining the number of BHs in clusters relative to X-ray observations of accreting BH binaries or gravitational wave observations. Once ejected from a dense GC, a BH-MS binary can be studied further using radial velocity observations, whereas this is not typically the case when inside a dense crowded cluster since radial velocity surveys cannot be performed due to crowding. This approach will allow us to verify the properties of the identified binary systems and potentially uncover systems that may contain BHs.

Another application of our method for detecting RS-RB pairs is in the study of high-velocity stars, many of which are observed in the Galaxy but do not have trajectories that point back toward the Galactic center. One hypothesis to explain these stars is that they originate from three-body ejection events in GCs. By analyzing the frequency of these interactions, we can determine whether all such high-velocity stars are consistent with having come from dense GCs, as opposed to the Galactic center. In future work, we aim to search for BH-MS binaries ejected by single MS stars. Even without population statistics, we can compare the observed properties of these binaries with theoretical distributions from three-body interactions to assess their rarity. By including more GCs in our sample, we can further strengthen the connection between theory and observation and address the origin of all high-velocity stars in the Galaxy.

In Sect. ~\ref{3body}, we present and discuss the relevant physical concepts needed to understand how and why our selection criteria are constructed and the software, \texttt{Corespray}, used to compute the predicted outcome distributions. In Sect. \ref{Methods}, we present our methodology for identifying the most probable RS-RB pairs.  In Sect. \ref{Results}, we present the results of applying our method to the old OC M67 using the Gaia data and identify the most probable pairs of RSs-RBs for further study (e.g., radial-velocity observations).  Finally, in Sect. \ref{Discussion}, we conclude and discuss improvements on the present analysis that can be implemented when applying our method to other clusters.

\section{Theoretical expectations for three-body disintegrations} \label{3body}

In this section we present the basic theoretical expectations for three-body disintegrations that go into deciding how our final selection criteria are constructed, in addition to the software \texttt{Corespray} that we use to obtain predicted outcome distributions from theory (e.g.,  ejection velocities, directions, etc.).  \texttt{Corespray} provides the theoretical information we need to be able to directly compare our observational results to theory.

\subsection{Causality and conservation of momentum and energy}

In this section we discuss in more detail the expected observed properties of RS-RB pairs ejected from clusters due to single-binary interactions and the underlying physical mechanisms that produce their properties.  We show that for the triple disintegration scenario, a larger number of selection criteria can be applied to the data to constrain the origins of RS-RB pairs relative to scenarios where only a single RS is expected to be produced (e.g., the SN-based mechanism).

To quantify what we expect for typical velocities of RSs formed from the triple disintegration scenario, we can compute a theoretically expected distribution of escaper velocities. Let the velocity of the escaper in the center-of-mass coordinate system be $v_{\rm{s}}$.  Then, a typical escape velocity distribution for the single star in three-body disintegrations can be described by the formula \citep{valtonen_karttunen_2006}

\begin{equation} \label{eqn:vel}
    f(v_{\rm{s}})dv_{\rm{s}} = \frac{(3.5|E_{\rm{0}}|^{7/2}m_{\rm{s}}M/m_{\rm{B}})v_{\rm{s}}dv_{\rm{s}}}{(|E_{\rm{0}}|+\frac{1}{2}(m_{\rm{s}}M/m_{\rm{B}})v_{\rm{s}}^2)^{9/2}},
\end{equation}

\noindent which is suitable to the isotropic case.  Here, the total energy of the three-body system is

\begin{equation}
    E_{\rm{0}} = \frac{1}{2}m\dot{r}^2_{\rm{s}} + \frac{1}{2}\mathcal{M}\dot{r}^2 - G\frac{m_{\rm{a}}m_{\rm{b}}}{r}-G\frac{m_{\rm{s}}m_{\rm{B}}}{r_{\rm{s}}},
\end{equation}

\noindent where the separation between the components of the binary system is $r$ and $r_{\rm{s}}$ is the separation between the escaper and the binary center of mass.  For the masses of the bodies we have: $m_{\rm{a}}$ and $m_{\rm{b}}$ are the masses of the stars in the binary system, $m_{\rm{B}}$ is the total binary mass ($m_{\rm{a}} + m_{\rm{b}}$), $m_{\rm{s}}$ is the mass of the single star escaper and $M$ is the sum of all masses in the three-body system. We also have $\mathcal{M} = \frac{m_{\rm{a}}m_{\rm{b}}}{M}$ and $m = \frac{m_{\rm{B}}m_{\rm{s}}}{M}$, which correspond to the reduced masses of the relative motions of the binary and the third body, respectively. The peak of this distribution is obtained by differentiating Eq.~\ref{eqn:vel} and setting it equal to zero: 

\begin{equation}
\label{Val_Car_Vspeak}
    (v_{\rm{s}})_{\rm{peak}} = \sqrt{\frac{2(M-m_{\rm{s}})}{5m_{\rm{s}}M}}\sqrt{|E_{\rm{0}}|}.
\end{equation}

The most likely escape velocity assuming an initial binary separation of 100 AU (integrating over \"Opik's Law from twice the orbital separation corresponding to a contact state to twice the hard-soft boundary yields an average orbital separation of $\sim$ 168 AU) is of the order of a few to a few tens of kilometers per second.  We note, however, that deriving a predicted velocity distribution requires assuming an underlying distribution of binary orbital parameters, both for the triple disintegration scenario as well as for the SN-based mechanism.

Given an assumed ejection velocity for an RS candidate coming from the triple disintegration mechanism, both timescales as well as momentum conservation-based arguments can be used to predict the exact position and velocity of the associated RB candidate.  In particular, since momentum conservation should be upheld during the decay of triple systems, the ratio of the RS and RB velocities should be equal to the inverse of the ratio of their masses, or

\begin{equation}
\label{eqn:momconserv}
v_{\rm RB} = \frac{m_{\rm RS}}{m_{\rm RB}}v_{\rm RS},
\end{equation}
where $v_{\rm RS}$/$v_{\rm RB}$ and $m_{\rm RS}$/$m_{\rm RB}$ are, respectively, the RS-RB velocities and masses.  Given a mass-luminosity ratio, the masses can be converted to observable luminosities (or magnitudes).  Additionally, the candidate RS and its associated candidate RB should have equal travel times from the position where the velocity vectors intersect.  We call this position the cross-point, and it should lie within the tidal radius of the host cluster.  Finally, using energy conservation, it is in principle possible to also constrain the most likely initial binary orbital separation of the putative single-binary interaction, given observational constraints on the final binary orbital separation as well as all three stellar masses.  We emphasize that none of the preceding constraints can be applied to RSs coming from the SN-based mechanism (or any other mechanism producing only a single RS) because they do not have an associated RB, which should in principle translate into an ability to more confidently associate an identified RS-RB pair candidate with a host (old) star cluster for the triple disintegration mechanism.

Few constraints on the orbital properties of the underlying binary population are available, particularly in old, massive clusters like GCs with crowded fields of view. We argue in this paper that by looking observationally for the products of single-binary interactions, limits can be placed on the distribution of binary orbital separations.  It is possible to obtain these constraints because the mean timescale between single-binary interactions in star cluster cores (where most ejection events are thought to occur) is given by \citep{Leigh_2011}

\begin{align}
\label{eqn:tau12}
\tau_{1+2} = 3.4 \times 10^7 (1-f_b)^{-1}f_b^{-1} \Big( \frac{ 1 {\rm \: pc}}{r_c} \Big)^3 \Big( \frac{10^3  {\rm \:pc}^{-3}}{n_0} \Big)^2 \\ 
\Big( \frac{v_{\rm rms}}{5 {\rm \: km} {\rm \: s}^{-1}} \Big) \Big( \frac{m}{1 {\rm M}_{\odot}} \Big) \Big( \frac{1 {\rm AU}}{a} \Big) {\rm yr},\nonumber
\end{align}
where $f_b$ is the binary fraction, $r_c$ is the core radius in parsecs, $n_0$ is the number density in the core in pc$^{-3}$, $v_{rms}$ is the root-mean-square velocity in kilometers per second, $m$ is the average mass (in M$_{\odot}$) and $a$ is the mean binary semi-major axis (in AU).  We note that Eq.~\ref{eqn:tau12} suggests that the rate of single-binary interactions scales linearly with the binary semi-major axis, which is the most difficult quantity in this equation to measure observationally.  It follows that, by constraining the frequency of RSs formed from single-binary interactions, limits can be placed on the underlying binary orbital separation distribution.  Based purely on Eq.~\ref{eqn:tau12}, it is specifically the mean semi-major axis that is constrained.

\subsection{Corespray}
\label{sec: corespray}
To obtain more robust predictions, including outcome distributions for disintegrating triples, we use \texttt{Corespray\footnote{For a complete description of \texttt{corespray}, please visit \url{https://github.com/webbjj/corespray}.}} to generate a distribution of RS-RB pair escaper velocities, as a function of our assumptions for the properties of the underlying binary population. \texttt{Corespray} is a PYTHON-based three-body particle spray code that simulates extra-tidal stars ejected from a star cluster's core due to single-binary interactions \citep{Steffani_2023_a}. All orbits in
\texttt{corespray} are defined and integrated using \texttt{galpy}\footnote{For a complete description of galpy, please visit \url{http://github.com/jobovy/galpy}.} a Python-package for galactic dynamics \citep{Bovy_2015}. We use KingPotential and MWPotential2014 from \texttt{galpy} to encompass influences from both the cluster and the galaxy. \texttt{Corespray} uses the formalism from \cite{valtonen_karttunen_2006} to generate the outcome distributions from three-body disintegrations.  The code allows us to generate a large sample of ejected RS-RB pairs, from which we obtain the expected distribution of ejection velocities.  

\begin{figure}
\centering
\includegraphics[width=\linewidth]{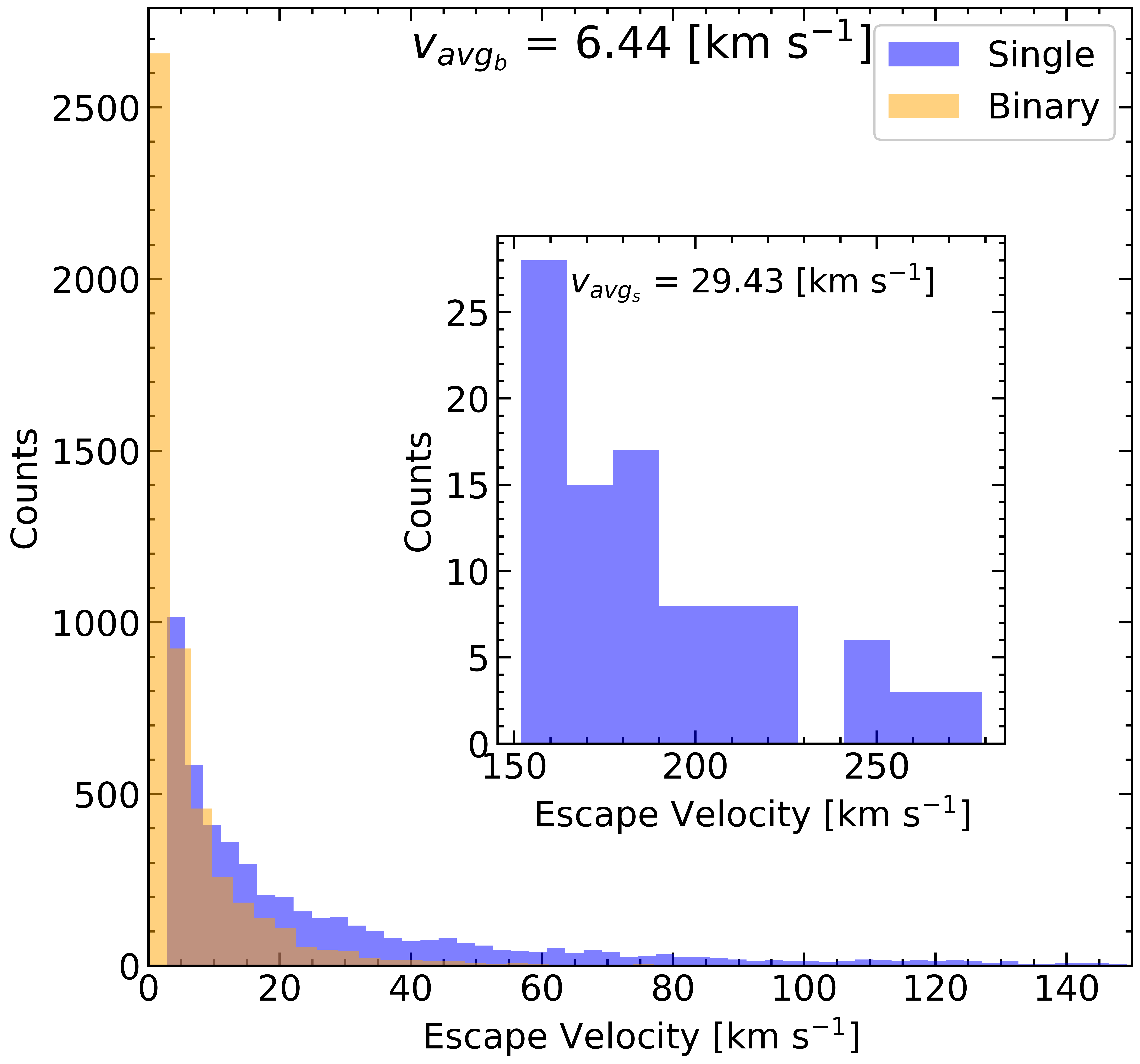}
    \caption{Histogram showing the predicted velocity distributions of escapers ejected in the \texttt{Corespray} simulation. The x-axis shows the ejection velocity in kilometers per second.}
    \label{fig:vel_corespray}
\end{figure}

We simulate with \texttt{Corespray} a sample of 5000 pairs ejected randomly over a period of 10 Myr. The cluster M67 is modeled with a King profile for the gravitational potential \citep{king_1962} and positioned in the Galaxy such that its center of mass and center-of-mass motion are in agreement with observational constraints (see Sect.~\ref{preliminary} for the exact values). We assume a range of stellar masses between 0.1 M$_{\odot}$ - 1.4 M$_{\odot}$ adopting a Salpeter mass function (i.e., with a power-law of $\alpha = -1.35$; \citealt{Salpeter_1955}). For the minimum and maximum orbital separations, or $a_{min}$ and $a_{max}$ respectively, we assume limits between twice the orbital separation corresponding to a contact state and twice the hard-soft boundary (assuming two solar mass stars). Using the data from \texttt{Corespray} we obtain the expected velocity distribution for the ejected pairs, as shown in Fig.~\ref{fig:vel_corespray}.

\begin{figure}
\centering
\includegraphics[width=\linewidth]{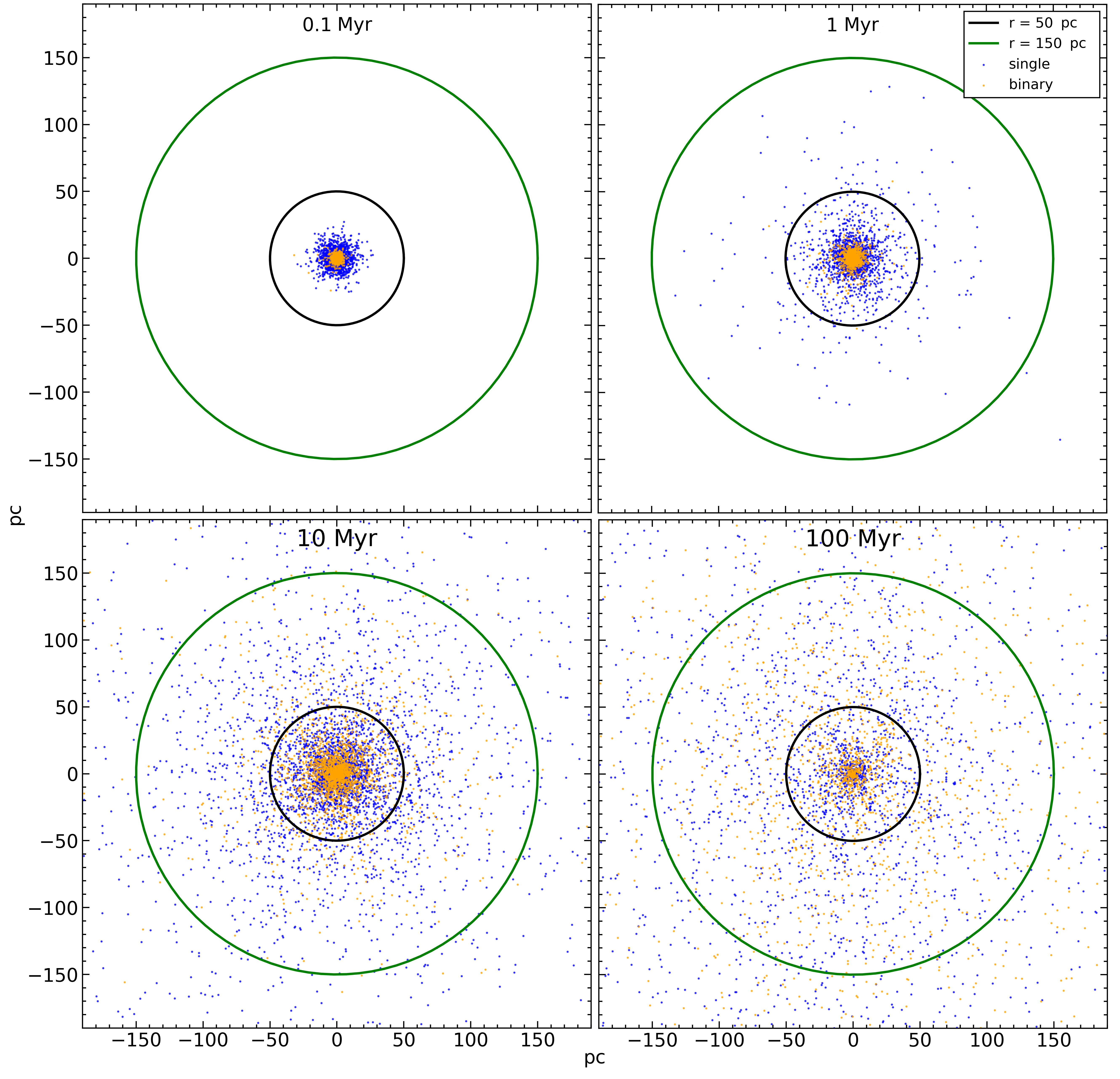}
    \caption{Four different snapshots in time of a simulation using \texttt{Corespray}. We show the spatial distribution of the ejected pairs in the plane of the sky at 0.1, 1, 10, and 100 Myr.}
    \label{fig:4panelplot}
\end{figure}

Next, we consider how many RSs we expect to currently observe around M67 due to triple disintegration coming from single-binary interactions in the cluster core.  Specifically, consider a circular field of view of a radius of 150 pc at the distance of M67. We chose this value for the radius initially to be a balance between computational expense (which increases with the number of candidate pairs) and the probability of finding an ejected RS-RB pair in the field of view.  Moving at a tangential velocity of 30 km s$^{-1}$ (where \texttt{Corespray} predicts an average velocity of 29.4 km s$^{-1}$ for RSs and 6.44 km s$^{-1}$ for RBs; see Fig. \ref{fig:vel_corespray}), a star will cross this field-of-view in roughly 10 Myr, as shown from our \texttt{Corespray} simulations in the bottom right panel of Fig. \ref{fig:4panelplot}.  But the time between single-binary interactions in M67 is $\lesssim$ 5 Myr. These timescales are not directly correlated; however, comparing them reveals that, theoretically, the detection of one or more pairs is indeed expected. This comparison highlights the consistency between the theoretical framework and the anticipated observational outcomes, reinforcing the plausibility of such detections. To calculate the mean time between interactions, we used the same parameters as adopted in Sect. 3.1 of \citet{Leigh13}. We adopted \"Opik's Law \citep{Opik_1924} for the initial orbital parameter distribution assumed in \texttt{Corespray}, and we integrated over this from twice the separation corresponding to a contact state to twice the hard-soft boundary yields an average orbital separation of $\sim$ 168 AU.  
Assuming all of these interactions resulted in an ejection velocity for the single star of $\sim$ 30 km s$^{-1}$ by a binary with a semi-major axis of 100 AU (i.e., close to the mean value of 168 AU for our assumed initial semi-major axis distribution), we therefore expected to observe on the order of two RS-RB pairs originating from the cluster in this field of view at the present time due solely to single-binary interactions. An observation of more or less than this amount could imply that some of our assumptions for our timescale calculation are incorrect, such as the mean binary orbital separation and/or the binary fraction.  Based on the above calculation for the expected number of pairs in our chosen field of view, we expect few RS-RB pairs to come from individual clusters, but this limitation on statistical analyses can be somewhat mitigated by including other clusters in the sample.

To summarize this section, given the additional constraints available when searching for RS candidates coming from the triple disintegration scenario, we intuitively expect to be able to more confidently associate them with their host cluster relative to runaway-producing ejection mechanisms that produce only a single RS. Hence, the inclusion of the RB in the analysis increases the number of selection criteria we have to work with.

\section{Methods}\label{Methods}

In this section we develop a methodology to identify RS candidates from clusters coming from the disintegration of three-body systems formed from single-binary interactions, which we apply to the old OC M67 as a case study. To this end, we present our methodology for identifying not only candidate RSs, but also candidate RBs that are consistent with having interacted with them in M67. The methodology results in a (short) list of RS-RB pair candidates and their corresponding observed properties.

\subsection{Data acquisition}

Here we introduce the data used to search for RS-RB pairs in the vicinity of M67, as well as our preliminary cuts in distance and position on the sky (but defer more in-depth calculations to Appendix \ref{appendix}).

\subsubsection{Gaia DR3 data}

Ideally, we must build a 6D phase space, requiring right ascension, declination, 2D proper motions, radial velocity and parallax (i.e., distance). In our case, we are able to construct a nearly complete 5D phase space using the Gaia DR3 catalog. 

Excluding radial velocities quantifies how constraining the data are without the third dimension in velocity, which can be exceedingly difficult to obtain for many, especially more distant, clusters and especially for those with high densities and hence crowded fields of view, such as GCs. It is also important to note that the Gaia parallaxes are somewhat limited in providing a precise distance for distant clusters like M67.  The median parallax uncertainties are 0.02-0.03 mas for G < 15, 0.07 mas at G = 17, 0.5 mas at G = 20, and 1.3 mas at G = 21 mag \footnote{\url{https://www.cosmos.esa.int/web/gaia/dr3}}. Hence, we adopted a course-grained cut on distance for our candidates, as described in the subsequent section.

This data release represents a major advance with respect to Gaia DR2 and Gaia EDR3 because of the unprecedented quantity, quality, and variety of astrophysical data. Even though Gaia DR3 includes the sixth data release from the Radial Velocity Experiment (RAVE) survey \citep{Steinmetz_2020}, we do not include these data in our analysis. From the total number of sources in Gaia only $\sim$ 1.8\% have a radial velocity value \footnote{\url{https://www.cosmos.esa.int/web/gaia/dr3}}.

\subsubsection{Preliminary sample selection} \label{preliminary}

In order to identify our initial sample of candidate RSs, we extract from the Gaia data \citep{gaiadr3_2023} a complete 5D kinematic data set (i.e., proper motions, parallax and position on the sky) that covers our chosen field of view centered on the motion of the center of M67 (as reported by \citealt{Jacobson_2011}) with a projected radius of 150 pc (i.e., forming a circular field of view in 2D). This limit and hence our first cut in the data is decided, as calculated in Sect.~\ref{3body}, based on the computed time between single-binary interactions weighed against the added computational expense of increasing our field of view and hence the total number of candidate pairs.  In the end, we expect order unity RS-RB pairs to be observed within our chosen field of view at any given time.

First, we apply a cut in distance.  For this, we adopt a cylindrical volume with a diameter and a length of 300 pc (in the direction of the cluster), centered on M67.  We select all stars that are inside this 3D cylinder.  We do not incorporate the uncertainties in parallax for this preliminary cut, which can be substantial in some cases.

Within this cylinder, we identify 203,190 objects not factoring in the uncertainties in parallax. Including objects with parallax uncertainties that make them consistent with lying within the domain of our cylinder to within 1$\sigma$ (one standard deviation) increases this number by about a factor of two.  We do not include these additional objects in our initial sample since most of them are dim with large uncertainties in much of our 5D parameter space. It is the dimmest stars that tend to have the largest uncertainties, making it more likely that they will satisfy one of our criteria at the 1$\sigma$ level.  It is therefore arguably a better choice to omit these additional sources since we do not expect it to be possible in many cases to securely identify them as disrupted triples.   
We note that the change in sample size we obtain upon performing a similar cut in the 2D position on the sky (i.e., including all sources that lie within 1$\sigma$ of falling inside our circular field of view) is negligible given the small uncertainties. Hence, we do not include these additional objects in our initial sample either.

For our second cut in the data, we do not consider objects located within the cluster in projection on the plane of the sky.  Using for the cluster center $\alpha = 8^h51^m23^s.3$, $\beta = +11^{\circ}49'02"$ (J2000) \citep{Jacobson_2011}, we identify those RSs located outside the cluster tidal radius.  In previous studies, the tidal radius was found to be either $14\pm 2.5$ pc \citep{Keenan_1973}, 16.8 pc \citep{Davenport_2010} or 17.98 pc \citep{Kharchenko}. For simplicity and to choose a balance between these values, we adopt a tidal radius of 17 pc.  For all of our calculations, we assume the following additional parameters for M67.  We assume a total mass of the cluster $M_{cl} = 2100 \pm 600$ M$_{\odot}$ \citep{Geller_2015}, a core radius of $r_c = 1$ pc \citep{Bonatto_2003}, a half-light radius $r_h = 2.6$ pc \citep{Balaguer_2006} and a central 1D velocity dispersion of 0.59 $\pm$ 0.07 km s$^{-1}$ \citep{Geller_2015}.

Finally, we remove from our sample all objects with 2D velocities smaller than the escape velocity from M67 relative to the cluster center of mass. For this, we adopt an escape velocity of 2.6 km s$^{-1}$, calculated using the method adopted in \citet{Georgiev_2009} by assuming a correction factor of $f_{c,t}(1.2) = 0.09154$ (corresponding to the velocity once the object reaches the tidal radius). We also consider only sources for which tracing back the velocity vector passes within the tidal radius of M67.  This cut reduces our sample size from 203,190 to 15,346. As we show in Sect.~\ref{Discussion}, our chosen criteria are in some ways conservative since the inclusion of the third dimension in velocity (i.e., radial velocities) could yield velocity vectors that are inconsistent with having originated from M67.

\subsection{Final sample selection}

In this section we select from our preliminary sample to identify the most likely candidate RS-RB pairs that were ejected from M67 due to single-binary interactions.  We begin, as described in Sect.~\ref{3body}, with a total of 15,346 objects, yielding in total 117,864,981 unique combinations.

\subsubsection{Traceback position and time}

First, we identify every pair of objects whose 2D velocity vectors trace back to intersect within the tidal radius of M67.  This is done as follows.  Using Eq. \hyperref[A1]{A1}, we first transform to a coordinate system in velocity space that is centered on the motion of M67's center, as reported by \citet{Jacobson_2011}.  We then calculate in this frame the 2D proper-motion vector for every object in both magnitude and direction using the Gaia proper motions in both RA and Dec (\citealp{Leeuwen_2009}, \citealp{Gaia_2018}, \citealp{Getman_2019}, and \citealp{Kuhn_2019},  see Eqs. \hyperref[proper_motion_eq]{A2, A3, and A4}), along with the corresponding uncertainties.  

Using the proper motions and their uncertainties in RA and Dec, we calculate three vectors for each object. The "primary" vectors are defined by the actual parameters in position and proper motion reported in the catalog for every pair. For illustrative purposes, we show two additional vectors defined by the uncertainties in position and proper motion as shown in Fig. \ref{fig:method_image}, calculated using Eqs.~\ref{eqn:Angle_Equation}, ~\ref{eqn:Angle_Equation_Lower}, and ~\ref{eqn:Angle_Equation_Upper}, yielding maximum and minimum distances in this parameter space that separate these points of intersection from that of the primary.  With this, we then calculate three points of intersection, called cross-points, by applying the equations in Appendix \hyperref[Crosspoint Equation]{A3}. We keep in our sample all pairs with at least one of these three cross-points lying within the tidal radius of M67 and discard the rest.

Next, all pairs of candidate objects with cross-points located within the domain of M67 must coincide in not only space but also time. For a chosen cross-point that is located within the M67 tidal radius, we calculated the time expected for both sources to get to that point tracing back in time. Hence, we calculated for both objects in every pair the traceback time (TBT) to the location of the cross-point, defined as the time taken to travel from the cross-point to the presently observed object position. This is done using the measured values in both distance traveled and velocity, assuming the presently observed velocity is constant and has not changed over the extent of its travel across the field of view.  We then calculate an uncertainty by propagating the provided uncertainties in both position and velocity.  Hence, the time it will take object a (and it is the same for object b) to reach the cross-point is given by

\begin{align}
    TBT_{\rm{a}} &= \frac{|\vec{r}_{\rm{a}}-\vec{r}_{\rm{cp}}|}{|\vec{v}_{\rm{a}}|} = \sqrt{\frac{\left(r^{\rm{x}}_{\rm{a}}-r_{\rm{cp}}^{\rm{x}} \right)^2+\left(r^{\rm{y}}_{\rm{a}}-r_{\rm{cp}}^{\rm{y}} \right)^2}{\left(v^{\rm{x}}_{\rm{a}} \right)^2+\left(v^{\rm{y}}_{\rm{a}}\right)^2}}
\end{align}

\begin{align}
    TBT_{\rm{b}} &= \frac{|\vec{r}_{\rm{b}}-\vec{r}_{\rm{cp}}|}{|\vec{v}_{\rm{b}}|} = \sqrt{\frac{\left(r^{\rm{x}}_{\rm{b}}-r_{\rm{cp}}^{\rm{x}} \right)^2+\left(r^{\rm{y}}_{\rm{b}}-r_{\rm{cp}}^{\rm{y}} \right)^2}{\left(v^{\rm{x}}_{\rm{b}} \right)^2+\left(v^{\rm{y}}_{\rm{b}}\right)^2}},
\end{align}

\noindent where $\vec{r_{\rm{a}}}$ is the distance separating the object's current position from the center of mass of M67 and $\vec{v_{\rm{a}}}$ is the velocity relative to the center-of-mass motion of M67.  Similarly, $\vec{r}_{\rm{cp}}$ is the vector from the center of mass of M67 to the cross-point with $r_{\rm{cp}}^{\rm{x}}$ and $r_{\rm{cp}}^{\rm{y}}$ being the corresponding values for each coordinate.  
If the difference in TBT values for a given pair of objects is consistent with zero to within 1$\sigma$, we keep the pair in our sample and discard it otherwise.

\subsubsection{Angle of intersection between velocity vectors}

\begin{figure}
    \centering
    \includegraphics[width=\columnwidth]{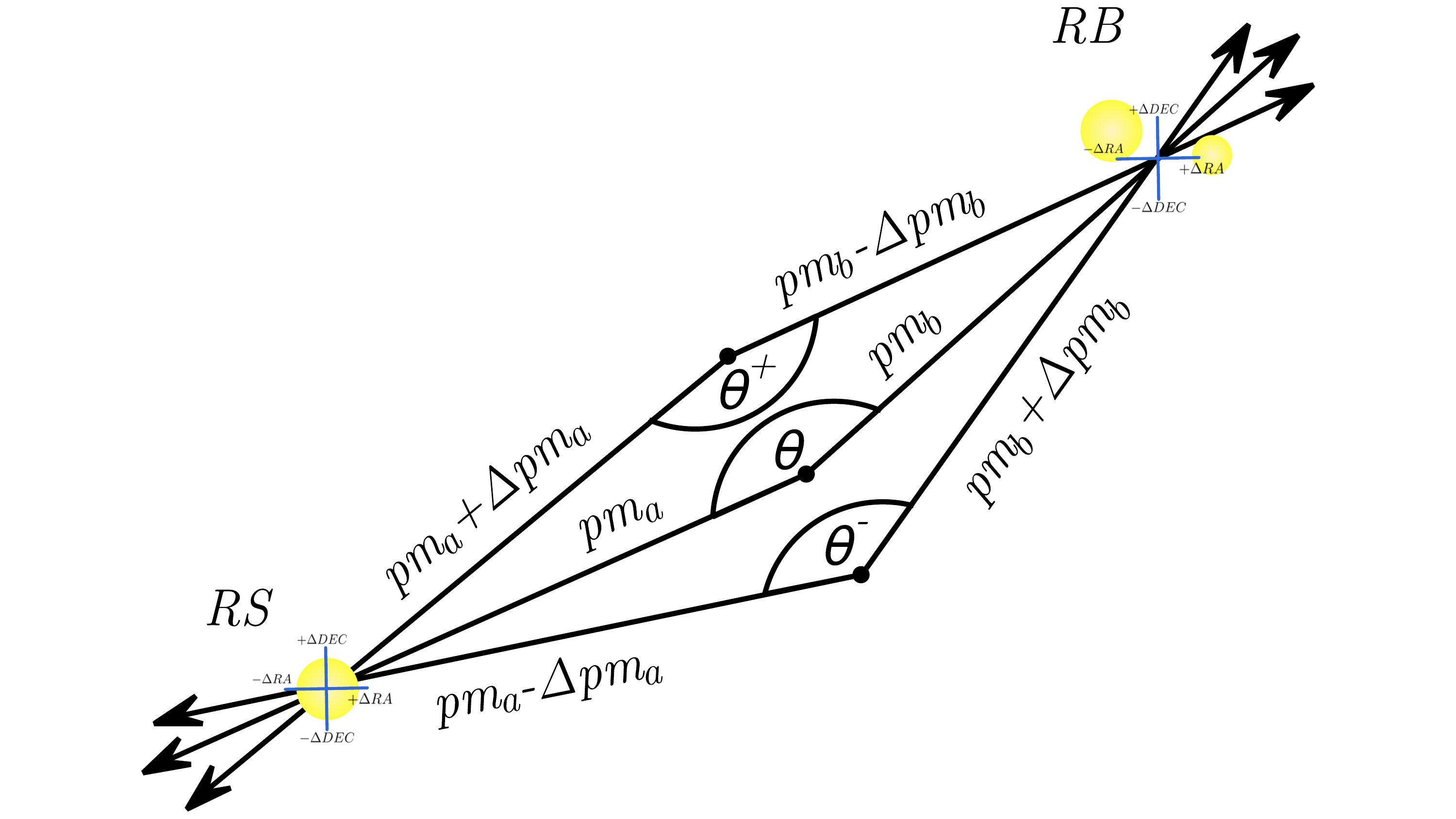} 
    \caption{Diagram illustrating the selection method for the angle of intersection.  We have the two limiting vectors described by the error in position and proper motion ($pm_{a} \pm \Delta pm_a$ and $pm_{b} \pm \Delta pm_b$) for the RS and RB, which yield a minimum (lower) and maximum (upper) possible angle (using the 1$\sigma$ uncertainties in both directions; see Eqs.~\ref{eqn:Angle_Equation},~\ref{eqn:Angle_Equation_Upper}, and ~\ref{eqn:Angle_Equation_Lower}). If a given pair has an angle of intersection consistent with 180$^{\circ}$ to within these upper and lower bounds, we keep it in our sample and discard it otherwise. We note that the RB is not resolved and the cross point is at the center of light.}
    \label{fig:method_image}
\end{figure}

We further required that in the plane of the sky, the angle between the velocity vectors (to the hypothetical cross-point) of a given pair must be consistent with 180$^{\circ}$, as illustrated in Fig.~\ref{fig:method_image}.  To evaluate this consistency, we used the two limiting vectors described above when evaluating the possible range of cross-points for each pair, which yield a minimum (lower) and maximum (upper) possible angle (using the 1$\sigma$ uncertainties in both directions; see Fig. \ref{fig:method_image}).   We calculate the angle between the intersection of each pair of vectors using Eqs.~\ref{eqn:Angle_Equation}, ~\ref{eqn:Angle_Equation_Upper}, and~\ref{eqn:Angle_Equation_Lower}. If a given pair has an angle of intersection consistent with 180$^{\circ}$ to within these upper and lower bounds, we keep it in our sample and discard it otherwise.

\subsubsection{Velocity ratio}

We also required that both objects have velocities consistent with linear momentum conservation.  This means that when assuming the slower moving object is a binary at least twice as massive as its single star counterpart (and that it was the lowest-mass star that was ejected from the three-body interaction as a single; \citealt{Heggie_1976}), the ratio of the object velocities should exceed two. We applied this cut to our data using the provided proper motions in RA and Dec by calculating a total magnitude for each object velocity in 2D (relative to the center-of-mass motion of M67), and then use these values to compute a ratio.  We propagate the 1$\sigma$ uncertainties in proper motions to compute an uncertainty for each velocity ratio.  We then keep in our sample all pairs that have a velocity ratio greater than two by 1$\sigma$ or more and discard them otherwise.

\subsubsection{Position in the color-magnitude diagram of the progenitor cluster}

Finally, we assess whether or not both objects in each pair have a location in the color-magnitude diagram (CMD) that is consistent with having originated from M67 using the Gaia DR3  $G_{BP}$ and $G_{RP}$ magnitudes.  Specifically, we required that one object (the candidate single star) lie within 1$\sigma$ of an isochrone for M67 in both color and brightness simultaneously (but we note that our results do not change significantly if we instead use both magnitudes instead of a color), while the other object (the candidate binary) lies within 1$\sigma$ (in both color and brightness) of the area in color-brightness-space defined as the region between the isochrone and an identical track shifted upward by 0.75 mag (corresponding to the equal-mass binary sequence).  

Additionally, it is the slower object that should be the brighter one from linear momentum conservation (assuming the brighter object is more massive) since the binary is most likely to include the most massive star as a dynamical outcome. To evaluate this, we use the PARSEC isochrone fit provided in \citet{Anna2023}. We consider that the pair is a valid candidate if the fastest object is the dimmest and it simultaneously lies within 1$\sigma$ of the isochrone in both magnitude and color.  At the same time, the slower object is the brightest and is consistent with lying between the isochrone and the equal-mass binary sequence to within 1$\sigma$ in both magnitude and color. If both objects in a given pair satisfy these criteria then we keep it in our sample and discard it otherwise.

To summarize, the criteria we implement are as follows, independent of the order in which they appear:

\begin{itemize}
    \item Traceback position: Identifies candidate pairs whose velocity vectors, rewound back in time, intersect within the cluster. Ensures a spatial coincidence.
    \item Traceback time: Checks if these candidates share the same travel time within errors to reach that intersection point. This approach ensures temporal coincidence.
    \item Angle of intersection: Angle between their velocity vectors, considering uncertainties, is consistent with 180 degrees in the plane of the sky.
    \item Velocity ratio: Only consider pairs with a velocity ratio exceeding two, with the brighter object moving more slowly.  The velocity ratio criterion is because it is the least massive object that is the most likely to be ejected as a single, and it should be moving faster which follows causally from linear momentum conservation given our assumptions.
    \item Position in the CMD: Examination of locations in a CMD and comparison with a model for the cluster. One object (the single star) should match the isochrone, while the other (the binary) should be brighter and lie within the region separating the isochrone from the equal-mass binary sequence.
\end{itemize}

\section{Results}\label{Results}

In this section, we first describe the results of our analysis, focusing on the fraction or number of pairs that simultaneously satisfy all of the criteria described in the previous section.  Our initial preliminary sample size, described in Section~\ref{preliminary}, is 117,742,185. 

In the end, we present the only pair that simultaneously satisfies all of our criteria, which is the pair with the smallest calculated uncertainties. The uncertainties in angle, TBT, and position in the CMD quickly become large when looking at the other eight candidates that pass all of our criteria (see Table~\ref{tab: table_1}).  

For each criterion, however, we evaluated to what extent it reduces our sample size and, wherever possible, compare to the results of our \texttt{Corespray} simulations to assess how each part of the criteria and our method performs when tested with simulated data.

\subsection{Traceback position and time distributions}

One of the cuts applied to the data ensures that the velocity vectors of a given pair of objects rewound in time (typically $\lesssim$ 1 Myr), intersect in 2D space within the tidal radius of M67. This cut alone reduces the sample size by 71\%.  We also required that these traceback vectors intersect not only in space but also in time, which represents a 90\% reduction. If the TBTs are consistent within the 1$\sigma$ uncertainties, then both objects in a given pair are taken as consistent with having left the cluster at the same time. The mean uncertainty on the TBTs is $\pm 0.201 Myr$. These cuts combined reduce our initial sample size by 97.46\%.  

\begin{figure*}
\centering
\includegraphics[width=1\linewidth]{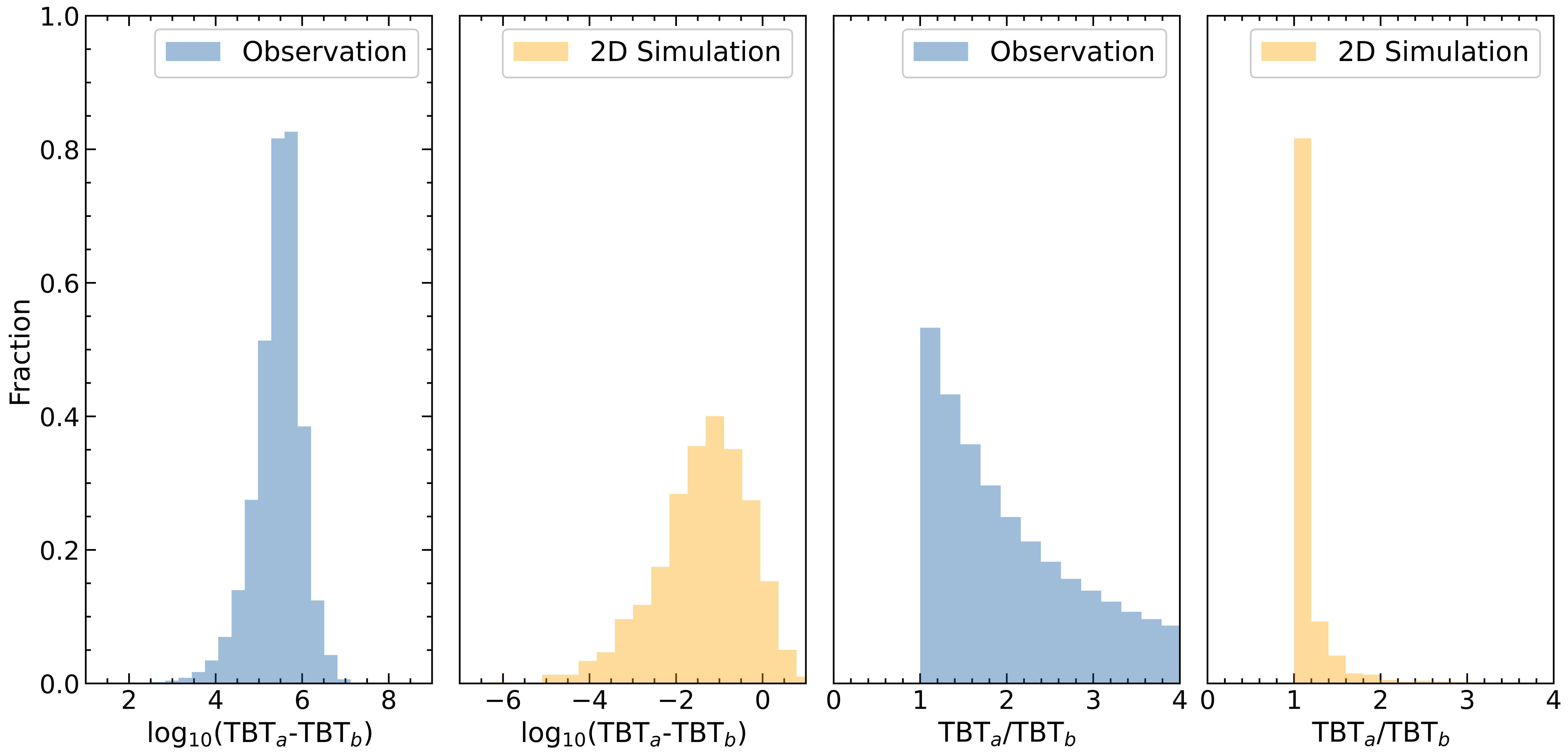}
    \caption{Histograms showing the distribution of the difference in TBTs (first and second panel) as well as their ratio (third and fourth panel) for every pair in our observational data and a 2D velocity projection in corespray.  The TBTs are provided in years.}
    \label{fig:TBT}
\end{figure*}

The distributions in both the ratio and differences and the ratios of TBTs for each pair are shown in Fig.~\ref{fig:TBT}. In Fig.~\ref{fig:TBT}, one can see that most TBTs are equal to within a factor of a few. We note that we do not compare to the \texttt{Corespray} results here since the difference in TBTs is always zero for these simulations.

The reason that these cuts only reduce our sample size by roughly an order of magnitude is that most objects have velocities that follow an overall Galactic flow and hence tend to point in similar directions with comparable magnitudes. Additionally, all objects lie a roughly similar distance away from the center of mass of M67 (i.e., to within a factor of a few given our field of view and the limiting tidal radius of M67).  Given these limitations, it is not so surprising that a relatively large fraction of pairs are consistent with having originated from M67 while also having similar TBTs.

\subsection{Distribution of angles of intersection between velocity vectors}

We further required that a given pair have an angle at the cross-point between the two velocity vectors that are consistent with 180$^{\circ}$, as predicted from linear momentum conservation.  To this end, we computed the angle of intersection for every pair of objects in our sample.  The resulting distribution is shown in Fig.~\ref{fig:angle}.

As is clear, the least probable angle observed is 180$^{\circ}$, making this cut especially constraining.  The typical uncertainty in this angle is highly dependent on how dim the two objects in a given pair are but is around a degree for pairs having both members brighter than $G_{bp} < 20$.  Hence, most sources do not fall within 1$\sigma$ of 180$^{\circ}$.  Specifically, this cut reduces our initial sample size by 99.78$\%$.

For comparison, we also show the predicted distribution of intersection angles we obtain from \texttt{Corespray} in Fig.~\ref{fig:angle}.\texttt{Corespray} predicts that the distribution should be much more clustered around 180$^{\circ}$, given that these pairs are the result of a three-body disintegration.  Hence, \texttt{Corespray} predicts that the more likely angle to be obtained is 180$^{\circ}$.

\subsection{Velocity ratio distribution}

In order to keep a given pair in our sample, we required that the velocity ratio of a given pair exceed two by at least 1$\sigma$.  The distribution of velocity ratios for every pair in our sample is shown in Fig.~\ref{fig:velocity ratio}, where the velocities are computed with respect to the cluster center-of-mass motion.  The vertical dashed line indicates the critical velocity ratio of two.  As is clear, this cut is not especially constraining.  It reduces our initial sample size by 69\%.

Curiously, most of the pairs in the \texttt{Corespray} results show a significantly broader velocity ratio than is observed, extending to much higher ratios. Most of the pairs in the \texttt{Corespray} simulation tend to have velocity ratios > 2, as theory predicts (given that \texttt{Corespray} only simulates the products of three-body disintegration). The surplus of velocity ratios > 2 relative to the observed distribution could be due to three reasons.  First, the effects of both the cluster and Galactic potential can significantly alter the ratio of velocities post-ejection, and \texttt{Corespray} adopts potentials that are not identical to the true underlying potentials.  Second, in order to properly compare theory to observations, the \texttt{Corepray} simulations would need to include foreground contaminants, which tend to have comparable velocities.  Third, \texttt{Corespray} provides 3D velocities which we compare to our 2D velocities projected onto the plane of the sky, and the third dimension can increase the velocity ratio.  We come back to the velocity ratio criterion in Section~\ref{galpotential}, but for now, we note that using our proposed velocity cut by choosing only sources with velocity ratios greater than 2 is likely not a useful criterion to incorporate in future work without a more sophisticated calculation that accounts for sources of acceleration/deceleration (e.g., cluster and/or Galactic potentials). However, removing this cut from our criteria for identifying the most probable pairs does not change our results significantly.

Finally, since we have assumed that the least massive object is always ejected from a three-body interaction, we also required that for each pair, the slower moving object be the brighter one. If we apply this simple cut after the one made previously in this section, we further reduce our sample size by one order of magnitude from the initial sample.  Hence, including a cut that combines the velocity ratio with the relative brightness of each object is much more constraining than a simple velocity cut alone.

\subsection{Position in the color-magnitude diagram}

Both objects should lie within 1$\sigma$ (simultaneously in both color and brightness) of the area in color-brightness-space defined by the region between the isochrone and an identical track shifted upward by 0.75 mag (corresponding to the equal-mass binary sequence).  Finally, we required that for a given pair, the fastest source (the candidate single) must lie within 1$\sigma$ (in both color and brightness) of an isochrone for the progenitor cluster. Further, considering the fact that the fastest star must lie on the isochrone to within 1$\sigma$ confidence, we eliminate approximately 98.54$\%$ of the total sample.

\begin{table}[h]
                \caption{Each criterion introduced and how it filters the total number of pairs from our catalog.}
\resizebox{\columnwidth}{!}{%
\begin{tabular}{|c|c|c|}
\hline
Filters & Sample Reduction & \begin{tabular}[c]{@{}c@{}}Pairs remaining\\ (117,742,185)\end{tabular} \\ \hline
Traceback space and time & 97.46\% & 2,990,651\\ \hline
Velocity ratio and brigthness & 90.44\% & 285,906 \\ \hline
Angle & 99.78\% & 628 \\ \hline
CMD & 98.54\% & 9\\ \hline
\end{tabular}%
}
\tablefoot{Considering each of the filters is independent we have a final number of the 9 resulting pairs. From this 9 remaining pairs 8 of them have considerable values of errors in angle, TBT and/or position in the CMD. Only one of them have reasonable errors and we consider it as our only result.}
\label{tab: table_1}
\end{table}

\subsection{Number of candidates remaining}
\label{sec: how many left}
\begin{figure}
    \centering
    \includegraphics[width=\columnwidth]{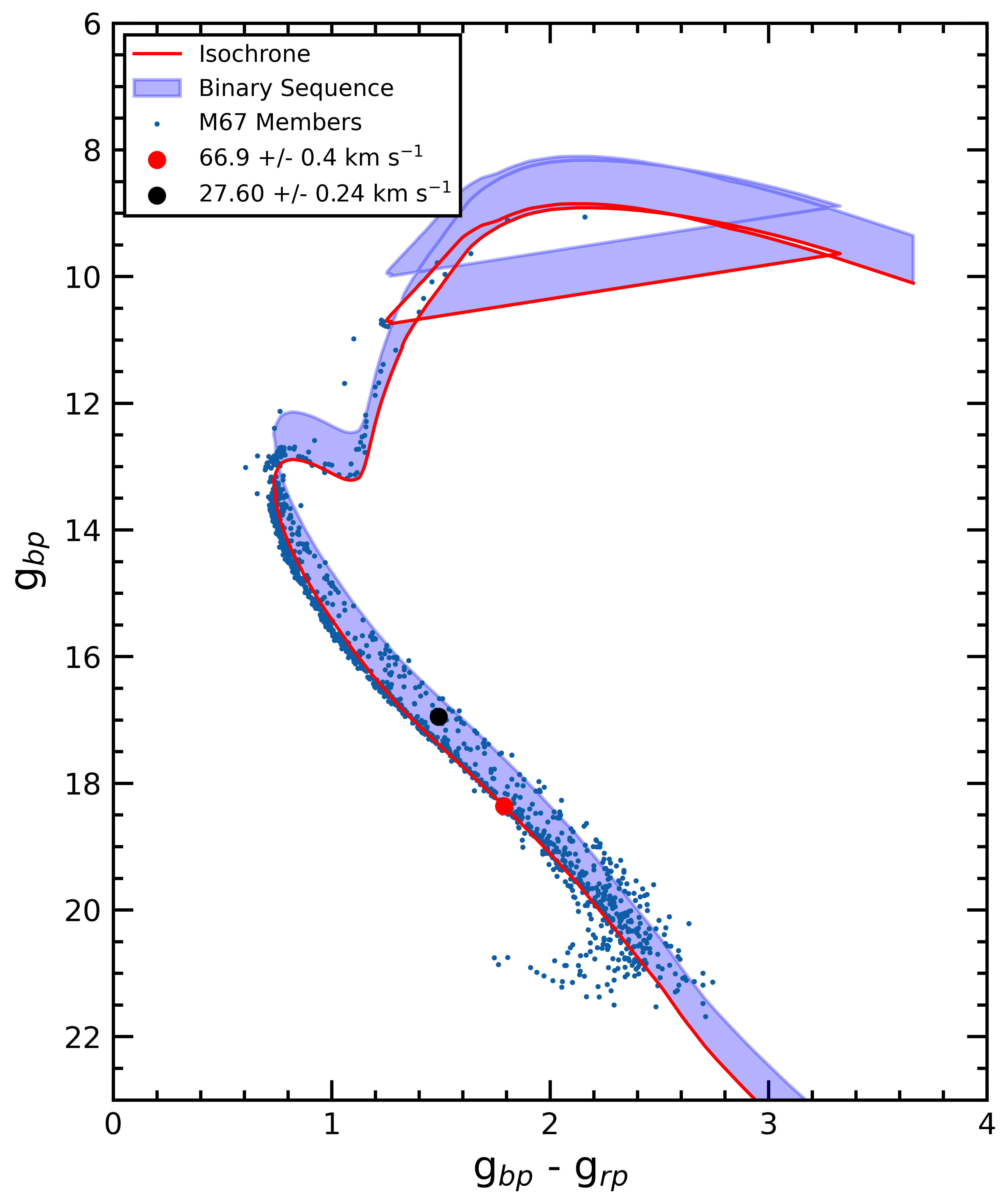} 
    \caption{Color-magnitude diagram of the old OC M67 plotted using the Gaia data. Only stars indicated as high-probability cluster members in \citet{Anna2023} have been selected and are plotted as small blue dots. The solid red lines show the best-fitting isochrone (lower line) as well as the equal-mass binary sequence shifted 0.75 mag above it (upper line). The larger dots indicate the two components of our one candidate pair that passes all of our criteria.  The red and black dots correspond to, respectively, the faster and slower moving objects. The cluster members do not include blue stragglers and other binary evolution products because they are far from a standard isochrone fit.}
    \label{fig:cmd}
\end{figure}

After applying all of the above criteria, we are left with only one pair that satisfies them all simultaneously. This one pair is shown in Fig.~\ref{fig:cmd}, and summarized below.  The small dots show high-probability cluster members obtained from \citet{Anna2023}, whereas the two large circles correspond to the two components of our one candidate pair.  The lines show an isochrone fit and the equal-mass binary sequence (obtained by shifting the isochrone up in brightness by 0.75 mag). The color coding indicated in the provided legend shows which of the two objects is the fastest, confirming that it is indeed the brighter object that is moving slower.

Therefore, after applying all of the criteria available to us from momentum conservation and causality, our total sample size is reduced by eight orders of magnitude, from $\sim$ 10$^8$ to 1.

This resulting pair has the following properties. For the RS, the properties are ID$_{a}$: 599619181604349440,
$v_a  = 66.9$ km s$^{-1}$, g$_{bp,a} = 18.36$, and g$_{rp,a} = 16.57$. And for the RB, they are ID$_{b}$: 604856670884640384, $v_b = 27.6 $km s$^{-1}$, g$_{bp,b} = 16.95$, and g$_{rp,b} = 15.46$. Between these two sources, we have the following: $v_{ratio} = 2.43$, $TBT_{diff} = 0.005 \pm 0.007$ [Myr], and $\theta_{a,b}^{\circ} = 179.6 \pm 0.5$. The RS candidate is consistent with lying on an isochrone in the CMD \ref{fig:cmd}, while the RB is consistent with lying within the binary sequence just above the MS.

A few additional properties of the RS-RB pair can also be inferred.  Given where the RS falls on the isochrone, we can infer that it has a mass of roughly 0.67 M$_{\odot}$. For the RB, we can think of the following limit. We assume that the vertical displacement of the binary from the isochrone could be explained if the primary component has a mass of 0.73 M$_{\odot}$ (i.e., as inferred from the vertical line connecting the RB's CMD location to the isochrone, a magnitude for the primary of 17.5).  We then took the magnitude required for the secondary in order to end up with a total magnitude equal to that of the binary, or M$_{\rm bin}$ $\sim$ 16.95.  Using this approach, we compute a magnitude for the secondary of 17.95, corresponding to a mass of 0.7 M$_{\odot}$ and hence a mass ratio of q $\sim$ m$_{\rm b}$/m$_{\rm a}$ $\sim$ 0.96, whereas before we assumed m$_{\rm a}$ and m$_{\rm b}$ are respectively the masses of the primary and secondary of the binary.   
Given these component masses, and the fact that the most likely semi-major axis is around 100 AU (see Section~\ref{sec: corespray}), the period of the binary would be 836.2 years. This corresponds to a low orbital velocity where detection with radial velocity is not feasible. However, at a distance of 1 kpc, 100 AU corresponds to 0.1 arcsec in the plane of the sky, which is resolvable from the ground.

\begin{figure}
\centering
\includegraphics[width=\linewidth]{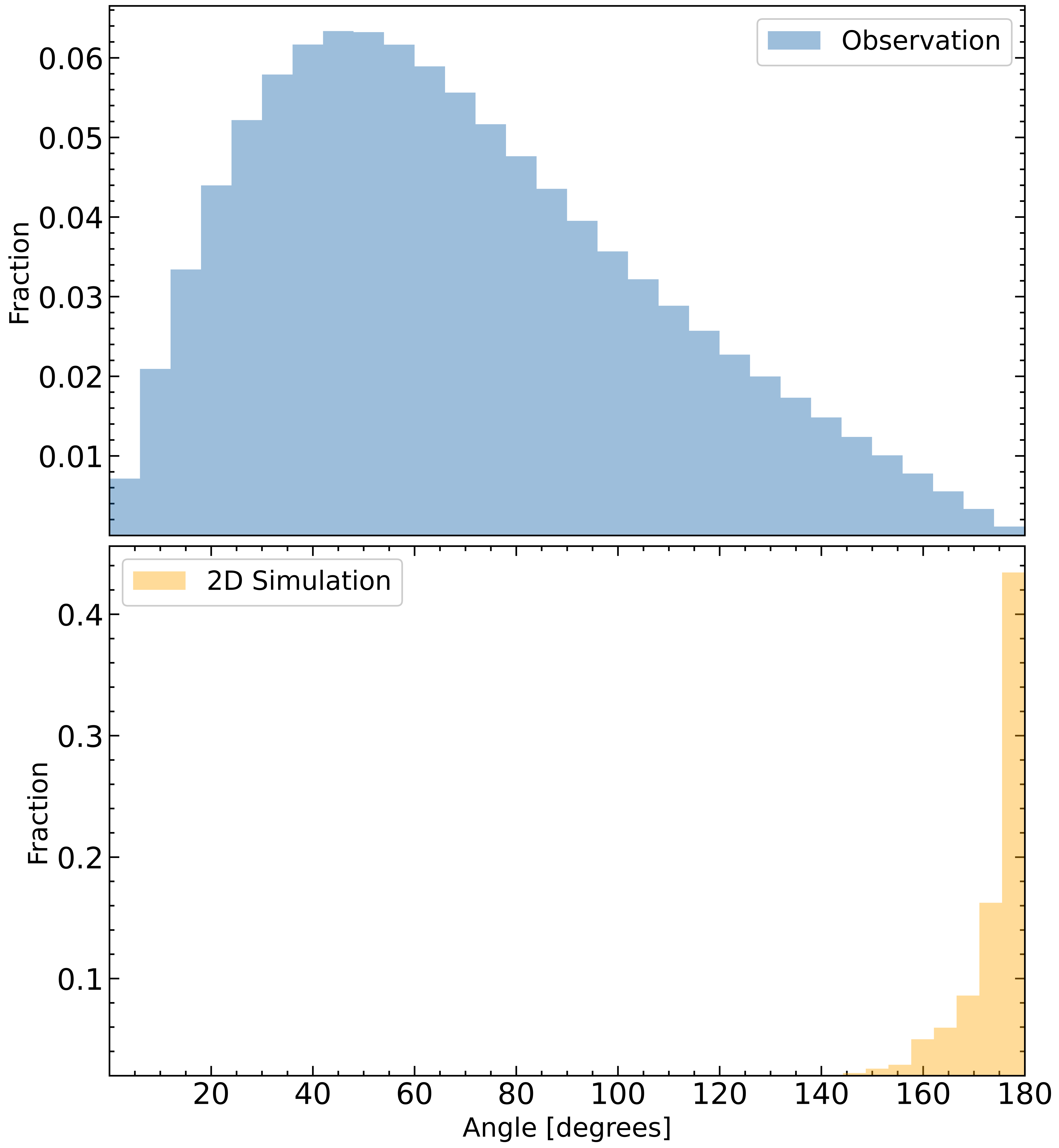}
    \caption{Histogram showing the distribution of intersection angles between the primary vector of every pair. The x-axis shows the angle in degrees. The top panel shows the projected angle for every pair from our \texttt{Corespray} simulations. The bottom panel shows the distribution with observational data. The simulation shows that most of the pairs tend to have angles near 180$^{\circ}$. In contrast, the pairs from the observational data set have a very low number of pairs near 180$^{\circ}$. This difference between the simulation (only pairs resulting from three-body disintegration) and observational (random sources in a field of view) distributions shows the angle to be one of the most important filters for the method.}
    \label{fig:angle}
\end{figure}

\begin{figure}
\centering
\includegraphics[width=\linewidth]{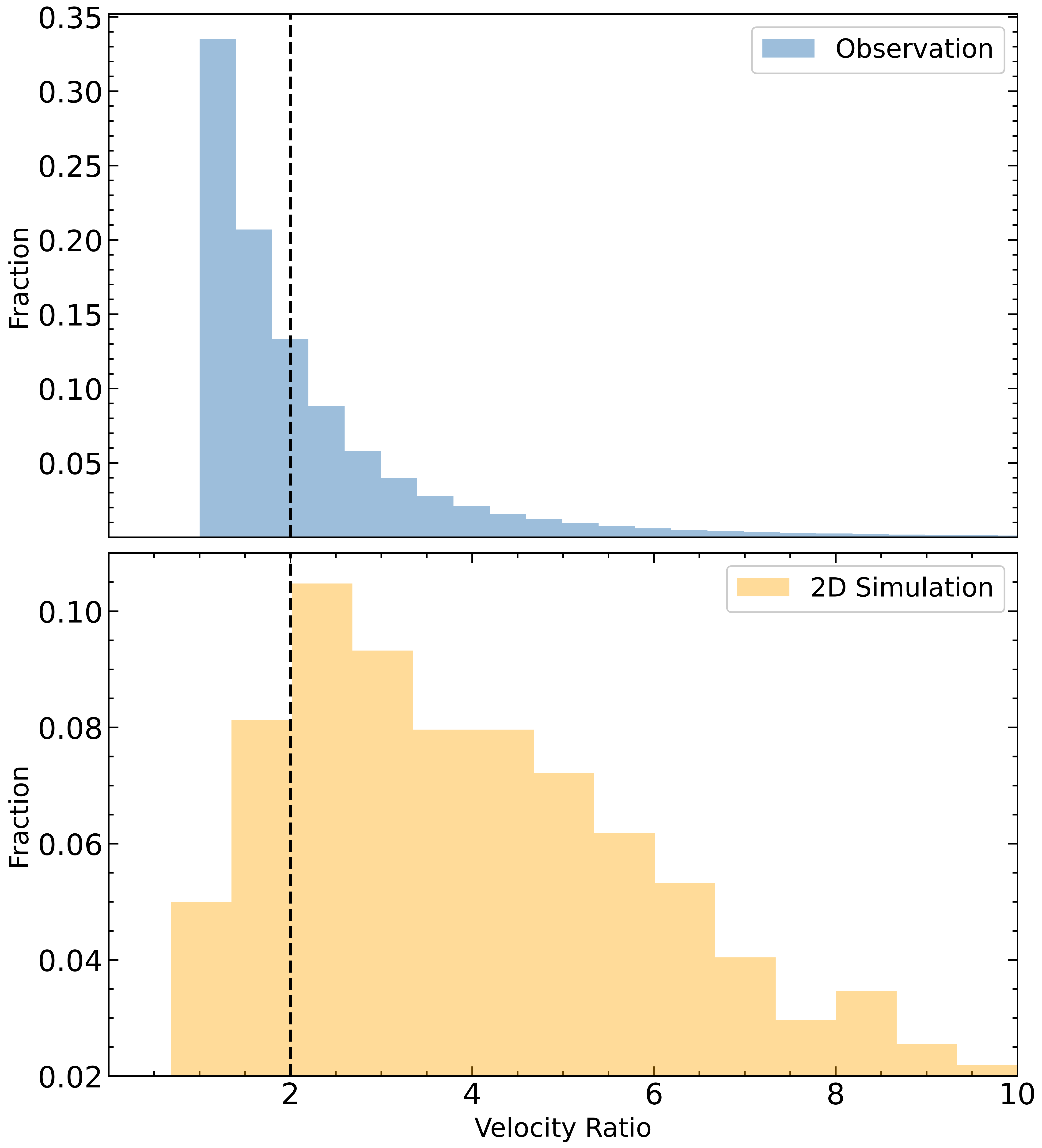}
    \caption{Histogram showing the distribution of velocity ratios. The top panel represents every pair from a simulation using \texttt{Corespray}, and the bottom panel shows all the pairs in our observational data. The vertical dashed line shows a critical ratio of two. Theory predicts that it is very likely that RS-RBs have velocity ratios greater than two, as can be seen in the simulated data set. On the other hand, the observational data show that more than half of the pairs have velocity ratios of less than two.}
    \label{fig:velocity ratio}
\end{figure}

\section{Discussion}\label{Discussion}

This section examines potential limitations of our methodology and suggests possible enhancements for future research. Addressing these areas could strengthen the approach and improve results.

\subsection{False positives}

How many false positives do we expect from our analysis?  Said another way, if we choose a random position on the sky for the cluster center-or-mass, how many pairs in a field of view comparable to the one assumed herein do we expect to simultaneously satisfy all of our selection criteria?  To address this, we choose a position on the sky shifted in either RA or Dec by 150 pc and then repeat our analysis for all pairs in this field of view.  This generates four different tests corresponding to the four different RA/Dec combinations (i.e., a positive/negative shift in RA and/or a positive/negative shift in Dec).  In total, this generates an initial sample of 468,586,715 pairs to which we apply our methodology, which is four times greater than the initial sample size we obtained.  Out of all these pairs, zero objects satisfy all of our criteria simultaneously (mostly due to the CMD location requirements).

Based on this analysis, we expect the frequency of false positive detection to be lower than 2 x 10$^{-9}$ and hence lower than the detection frequency found in this paper (i.e., 10$^-8$).  The most likely false positives should in general be associated with dimmer objects, due to the larger uncertainties. Despite having roughly four times more pairs compared to the original test, none passed all of our selection criteria.

\subsection{The effects of the galactic potential} \label{galpotential}

\begin{figure}
    \centering
    \includegraphics[width=\columnwidth]{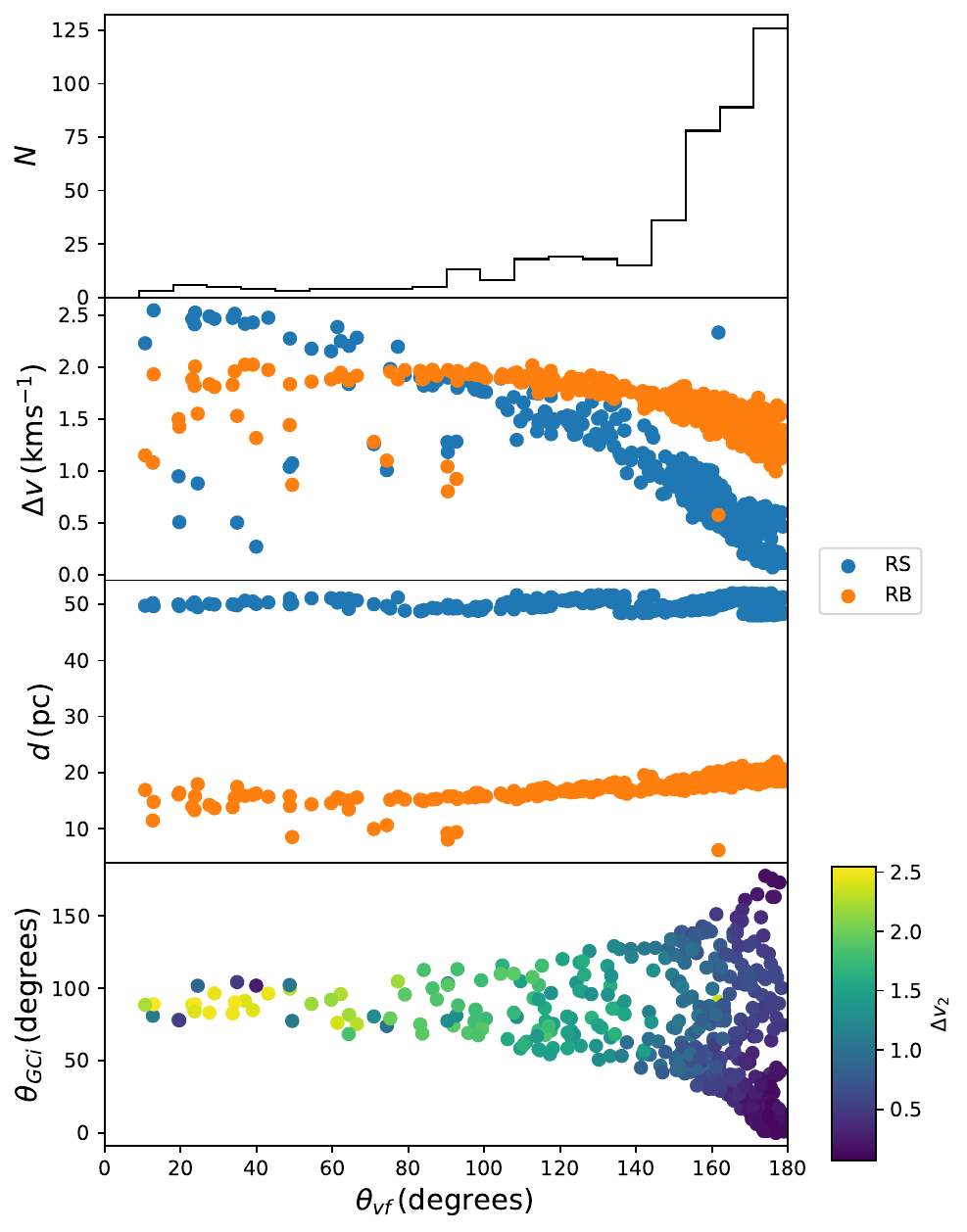} 
    \caption{Results of integrating the trajectories of escapers from M67 in the Galactic potential until the least massive object reaches 50 pc from the cluster center.  Each inset is plotted with respect to the final angle between the 3D velocity vectors of the escapers, $\theta_{vf}$. (The initial angle between the velocity vectors is 180$^{\circ}$).  The top panel shows a histogram of the distribution of $\theta_{vf}$ for all pairs.  The second panel shows the total change in velocity magnitude for each object in each pair.  The third panel shows the final distances from the cluster center. In the bottom panel, the y-axis shows the initial direction of ejection of the lighter star, where $\theta_{GC} = 0^{\circ}$ corresponds to ejection directly toward the Galactic center, and $\theta_{GC} = 90^{\circ}$ corresponds to ejection 
    in the direction of motion of the cluster.} 
    \label{fig:galpot}
\end{figure}

Our method neglects the effects of the cluster and Galactic potentials.  This is because the expectation is that any changes imparted to the velocity vectors of escapers from M67 should be small over our chosen field of view.  This assumption turns out to be a good approximation when both objects in a given pair have high velocities and are located close to the cluster such that their travel times to their currently observed locations are short.  However, this assumption can fail substantially when the object velocities are low and their positions are far from the host cluster.  This is illustrated in Fig.~\ref{fig:galpot}, which is meant as an independent calculation to what is shown from \texttt{Corespray} in Fig.~\ref{fig:velocity ratio} and Fig.~\ref{fig:angle}. Ultimately, Fig.~\ref{fig:galpot} confirms what was found in Figs.~\ref{fig:velocity ratio} and~\ref{fig:angle}, which suggests that velocity ratios significantly larger than 2 are common theoretically (see the top panel of Fig.~\ref{fig:velocity ratio}) but much less common observationally (see the bottom panel of Fig.~\ref{fig:velocity ratio}). 

Hence, a simple cut in only the velocity ratio need not be implemented in future work applying the method presented in this paper to other star clusters.  For now, we again note that removing this cut from our final sample selection does not affect our results.

Figure~\ref{fig:galpot} shows the results of simulating 500 ejection events from M67.  For each, the trajectories of the ejected objects are integrated into the Galactic potential using \texttt{AMUSE} (Astrophysical Multipurpose Software Environment, \citealp{Pelupessy_2013}), a component library for performing astrophysical simulations involving different physical domains and scales, and galpy: A Python Library for Galactic Dynamics \citep{Bovy_2015}. We work over a 50 pc distance (i.e., a distance representative of that for our identified candidate pair) from the center of mass of the progenitor cluster (for the single star only), and the final angle between their velocity vectors as well as the change in magnitude of the velocities are calculated.  The initial velocity of the binary is set equal to the cluster escape velocity, and the initial single star speed is then decided assuming linear momentum conservation.  Hence, we adopted the minimum possible ejection velocity, which maximizes the effects of the Galactic potential in changing the object velocities over our chosen field of view.  We note that we work with 3D velocities, but these results remain unchanged using 2D velocities.

As indicated by the color coding in the bottom panel of Fig.~\ref{fig:galpot}, most pairs experience the most significant change in velocity when ejected parallel to the orbital motion of the cluster and the smallest change in velocity when ejected perpendicular to it. We note as well that objects tend to be decelerated in the former case and accelerated in the latter.  As is clear, for those objects leaving the cluster with small escape speeds, it is indeed possible for the final angle of intersection between their velocity vectors to deviate significantly from 180$^{\circ}$.  This becomes even more so if a larger field of view is considered, and the orbits are integrated out to 150 pc or more.  If we instead adopt for the initial binary velocity ten times the cluster escape speed, then Fig.~\ref{fig:galpot} changes dramatically, and 100\% of the pairs have a final intersection angle within 5$^{\circ}$ of 180$^{\circ}$ since the typical uncertainty on this angle are a distribution around 4.4$^{\circ}$. We note that, for the one candidate pair identified in this paper, the observed velocities are of order a factor of ten larger than the cluster escape speed, and both objects are close to M67 in projection on the sky.  

\begin{figure*}
    \centering
    \includegraphics[width=1\textwidth]{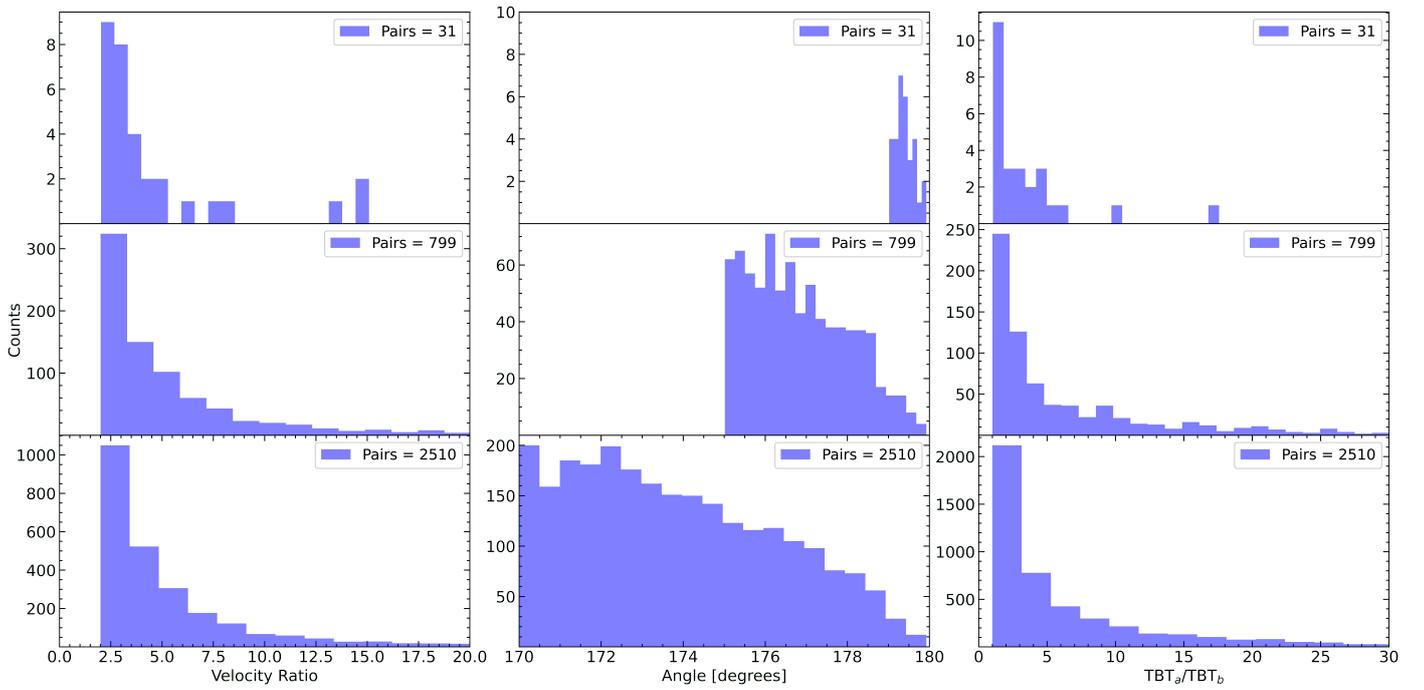}
    \caption{Resulting velocity ratio, angle of intersection, and TBT ratio distributions from \texttt{Corespray}. From top to bottom, we isolate pairs with angles greater than 179$^{\circ}$, 175$^{\circ}$, and 170$^{\circ}$ (but less than 180$^{\circ}$).}
    \label{fig:img1}
\end{figure*}

This general behavior is also found in our \texttt{Corespray} simulations.  As shown in Fig.~\ref{fig:img1}, including all sources with intersection angles $> 175^{\circ}$ (i.e., roughly 5$\sigma$ away from 180$^{\circ}$) reduces our sample size by less than a factor of two, whereas only including those sources with angles of intersection $> 179^{\circ}$ reduces our sample by more than two orders of magnitude.  Importantly, of those remaining 31 simulated pairs, only four have velocity vectors that are consistent with pointing orthogonally away from the direction of the orbital motion of the cluster to within 5$^{\circ}$.  The final sub-sample from \texttt{Corespray} then yields four pairs that satisfy all of our criteria while also having trajectories that should minimize the effects of the Galactic potential by traveling orthogonally to the orbital motion of the cluster (by $\pm 5^{\circ}$). 

Thus, \texttt{Corespray} predicts four pairs out of an initial sample of 5000 that meet the same criteria as our single best candidate, for which the observed parameters are summarized in Sect.~\ref{sec: how many left}. This potentially introduces another factor into our analysis, which suggests that many ejection events are likely needed before one will adhere to all of our selection criteria. How to calculate a correction factor coming from our theoretical analysis is non-trivial, and should be one of the focuses of future work.  But, the theoretical predictions from \texttt{Corespray} confirm that some candidate pairs adhering to our strictest selection criteria should exist, and typically be moving tangentially away from the cluster center-of-mass motion in the Galactic potential since this minimizes possible changes to the RS-RB velocity vectors post-ejection from the cluster. The importance of taking into account the Galactic field is highlighted by these results and ultimately summarized in Figs.~\ref{fig:galpot} and~\ref{fig:img1}:  the Galactic potential can erase any signature of a three-body decay significantly and quickly, and it becomes the dominant effect when within a few tidal radii from the host cluster.   

A calculation using conservation of energy suggests that a single star escaping the cluster from the central core will only have its velocity changed by less than 1 km s$^{-1}$.  Hence, we do not expect that taking the cluster potential into account will change our results significantly, suggesting as well that the Galactic potential is the more important effect, at least for such low-mass OCs as M67 and candidates with low velocities that lie far from the cluster tidal radius in projection.

To summarize, the results of this preliminary analysis suggest that the effects of the Galactic potentials should be considered in future work when searching for ejected RS-RB pairs.  The effect is minimized for faster objects located closer to the progenitor cluster, as is the case for our one resulting pair and confirmed by the \texttt{Corespray} and \texttt{AMUSE} simulations.

\subsection{The sixth dimension}

Unfortunately, a RAVE radial velocity is not provided for our candidate. In order to further. In order to further improve our method, radial-velocity follow-up observations can be used.  This cannot only constrain whether or not a given pair is consistent with having originated from the progenitor cluster along the line-of-sight, but also potentially the binary nature of the RB candidate and even its orbital properties.  
This makes additional predictions based on momentum conservation and causality that can be tested using radial velocity measurements (to confirm/reject that it is consistent with having originated from M67 and confirm/reject that it is a binary).

\section{Summary}

In this paper, we have presented a systematic methodology to identify RSs ejected from star clusters due to single-binary interactions. This was done by exploiting the fact that we expect such disintegrations to produce a pair of objects (i.e., a single star and a binary) whose relative velocity vectors, positions on the sky, and positions in the progenitor cluster CMD are consistent with linear momentum conservation and causality. In other words, both objects are consistent with having originated from within the cluster at the same time.  

Using the old OC M67 as a benchmark test case, we compared our theoretical expectations to a 5D kinematic data set obtained from the Gaia DR3 catalog via the provided parallaxes, proper motions, and positions on the sky. Our analysis yields, out of an initial sample size of roughly 10$^8$ candidate pairs, only one single-binary pair that simultaneously satisfies all of our selection criteria. We have presented the observed distributions for our entire sample to illustrate which of our criteria are the most constraining, and we compared these distributions to theoretical predictions, coming mostly from \texttt{Corespray}. We find that a small fraction (<1\%) of our initial sample of candidate pairs has an angle of intersection between their velocity vectors that is consistent with 180$^{\circ}$ to within 1$\sigma$.  Additionally, on the order of 10\% of our initial sample simultaneously has a ratio of velocities that exceed two, with the brighter of the two objects in the pair traveling the slowest.  Both of these criteria are expected from linear momentum conservation and tend to be more constraining than simple causality arguments related to the relative travel times and distances of both objects in a given pair being commensurate.  
With that said, the velocity ratio alone is not very constraining since velocity ratios greater than two are common (see the top panel in Fig. \ref{fig:velocity ratio}).

While the results obtained in this study for M67 are promising, it is important to bear in mind that our candidate pair is still preliminary. It should be further constrained using, for example, radial velocities and abundances as well as a more rigorous false positive test. Also, it is important to validate our method by applying it to other star clusters. This would not only further test the effectiveness of our method but would also potentially provide valuable insight into the dynamical evolution of other clusters. Furthermore, by studying multiple clusters, we could potentially identify trends in the properties of RS-RB pairs and test current theoretical predictions for the internal dynamical evolution of the progenitor clusters. That is, our method can reveal differences in runaway demographics between different clusters, which would give valuable information on the dynamical evolution of different types of clusters. In future work, we hope to apply our method to a large sample of clusters in order to generate the required statistics to address these questions.

\begin{acknowledgements}
AEHR acknowledges financial support from Millenium Nucleus NCN19\_058 (TITANs). NWCL gratefully acknowledges the generous support of a Fondecyt General grant 1230082, as well as support from Millenium Nucleus NCN2023\_002 (TITANs) and funding via the BASAL Centro de Excelencia en Astrofisica y Tecnologias Afines (CATA) grant PFB-06/2007.  NWCL also thanks support from ANID BASAL project ACE210002 and ANID BASAL projects ACE210002 and FB210003. 
AMG acknowledges financial support from NSF AAG grant no. AST-2107738. SMG acknowledges the support of
the Natural Sciences and Engineering Research Council of Canada
(NSERC) and is partially funded through a NSERC Postgraduate
Scholarship – Doctoral. RDM gratefully acknowledges financial support by the Fulbright U.S. Scholar Program, which is sponsored by the U.S. Department of State and the Chile-American Fulbright Commission. Its contents are solely the responsibility of the author and do not necessarily represent the official views of the sponsors.
\end{acknowledgements}

\section*{Data availability}

The data underlying this article will be shared on reasonable request to the corresponding author.

%
%

\bibliographystyle{aa}
\bibliography{example.bib}


\begin{appendix}
\section{Reference frame}
\label{appendix}
\label{reference_frame_eq}

\subsection{Positions}\label{positions_eq}

We begin with the positions for all the sources in the catalog, which we first consider in a two dimensional frame. Using $\mu_{\alpha *} \equiv  \mu_\alpha \cos\delta$, the coordinate transformations for the positions frame are 

\label{A1}
\begin{equation}
    \begin{split}
        x &= \cos{\delta}\sin{\left(\alpha-\alpha_{\rm{0}}\right)} \\
        y &= \sin{\delta}\cos{\delta_{\rm{0}}}-\cos{\delta}\sin{\delta_{\rm{0}}}\cos{\left(\alpha-\alpha_{\rm{0}}\right)} \\
    \end{split}
\end{equation}

The positions of the sources are denoted by $\alpha$  (RA) and $\delta$ (Dec). The center of the cluster is identified by its RA and Dec, taking values $\alpha_0 = 8^h51^m23^s.3$, $\delta_0 = +11^{\circ}49'02"$ (J2000) \citep{Jacobson_2011}

\subsection{Proper motions}\label{proper_motion_eq}
Our goal is to determine two components of stellar velocities based on the proper motions $\mu_{\alpha}$ and $\mu_{\delta}$.

The apparent motions of stars within stellar systems can be influenced by perspective effects and the motion of the system's center of mass, as noted in \cite{Leeuwen_2009}. In particular, the motion of a cluster in the radial direction can produce an effect known as "perspective expansion," which has been observed in Gaia measurements of GCs, many of which have radial velocities (RVs) reaching hundreds of kilometers per second \cite{gaia_2018_2}. A first-order approximation of perspective expansion can be derived from Eq. 13 in \cite{Leeuwen_2009}.
The equations for the additional shift in proper motion of a star are

\begin{equation}
    \begin{split}
        \Delta \mu_{\rm{\alpha *,per}} &\approx \Delta \alpha_{\rm{i}} \left( \mu_{\rm{\delta,0}} \sin{\delta_{\rm{0}}} - \frac{v_{\rm{r}}\varpi_{\rm{0}}}{\kappa}\cos{\delta_{\rm{0}}} \right), \\
        \Delta \mu_{\rm{\delta,per}}   &\approx -\Delta \alpha_i \mu_{\rm{\alpha *,0}} \sin{\delta_{\rm{0}}} - \Delta \delta_{\rm{i}}\frac{v_{\rm{r}}\varpi_{\rm{0}}}{\kappa}.
    \end{split}
\end{equation}

Here, $\Delta \alpha_i$ and $\Delta \delta_i$ are the differences in right ascension and declination between the system center of mass and an individual star.  The proper motions in RA and Dec are, respectively, $\mu_{\delta,0} = -10.97$ mas yr$^{-1}$ and $\mu_{\delta,0} = -2.9396 $ mas/yr$^{-1}$.  We assume $v_{\rm{r}} = 33.64$ $\pm$ 0.03 km s$^{-1}$ \citep{Geller_2015} for the radial velocity of the center-of-mass motion of M67 and $\varpi_{\rm{0}}$ = 860 pc for the cluster's parallax \citep{Cantat_2018}. The conversion factor that converts milliarcseconds per year to kilometers per second at a distance of 1 kpc is $\kappa = 4.74$. The first term in each equation accounts for motion within a spherical coordinate system, while the second term accounts for the perceived expansion or contraction of an object as its distance increases or decreases.

These contributions can be subtracted from the observed proper motions (relative to the system center of mass) to isolate the influence of internal kinematics \citep{Brown_1997}. We calculated the components of the corrected velocities parallel to lines of constant right ascension and declination:

\begin{equation}
    \begin{split}
        \mu_{\rm{\alpha,rest}}  &\approx - \kappa\left(\frac{\Delta \mu_{\rm{\alpha *,obs}}-\Delta \mu_{\rm{\alpha *,per}}}{\varpi_0}\right),\\
       \mu_{\rm{\delta,rest}}   &\approx \kappa\left(\frac{\Delta \mu_{\rm{\delta *,obs}}-\Delta \mu_{\rm{\delta *,per}}}{\varpi_0}\right)
    \end{split}
\end{equation}

with $\Delta \mu_{\rm{\alpha *,obs}} =  \mu_{\rm{\alpha *}}-\mu_{\rm{\alpha *,0}}$ and $\Delta \mu_{\rm{\delta *,obs}} = \mu_{\rm{\delta}}-\mu_{\rm{\delta,0}}$ are the differences in proper motion in right ascension and declination, respectively, between an individual star and the system center of mass. We have as a result the $\mu_{\alpha,\rm{rest}}$ and $\mu_{\delta,\rm{rest}}$ as the proper motions in the rest of frame calculated for RA and DEC. We then transformed the velocities to a 2D Cartesian coordinate system, $v_x$ and $v_y$, using the orthographic projection \citep{gaia_2018_2}:

\begin{equation}
  \begin{split}
    v_{\rm{x}} &= \mu_{\alpha,\rm{rest}}\cos{\left(\alpha-\alpha_{0}\right)}-\mu_{\delta,\rm{rest}}\sin{\delta}\sin{\left(\alpha-\alpha_{0}\right)} \\
    v_{\rm{y}} &= \mu_{\alpha,\rm{rest}}\sin{\delta_{0}}\sin{\left(\alpha-\alpha_{0}\right)} + \\
              & \mu_{\delta,\rm{rest}} \left( \cos{\delta}\cos{\delta_{0}} + \sin{\delta}\sin{\delta_{0}}\cos{\left(\alpha-\alpha_{0}\right)}\right).
  \end{split}
\end{equation}

\label{Crosspoint Equation}
\subsection{Cross-point and angle of the stars}\label{crosspoint_eq}

We used the following equation to find the cross-point. With $\Vec{X}$ equal to all the possible points on the line connecting each velocity vector to the center-of-mass motion of the cluster.:

\begin{equation}
    \Vec{X} = \lambda\hat{n} + \Vec{X}_{\rm{0}},
\end{equation}

The factor $\lambda$ determines the cross-point. The unit vector $\hat{n}$ that we adopt satisfies $\Vec{n} = \frac{\Vec{v}}{|\Vec{v}|}$. The vector equations of the line for sources a and b are:

\begin{align}
        \vec{X}_{\rm{a}} &= \lambda_{\rm{a}} \hat{n}_{\rm{a}} + \vec{r}_{\rm{a}} = \lambda_{\rm{a}} \frac{\vec{v}_{\rm{a}}}{|\vec{v}_{\rm{a}}|} + \vec{r}_{\rm{a}}\\
    \vec{X}_{\rm{b}} &= \lambda_{\rm{b}} \hat{n}_{\rm{b}} + \vec{r}_{\rm{b}} =\lambda_{\rm{b}} \frac{\vec{v}_{\rm{b}}}{|\vec{v}_{\rm{b}}|} + \vec{r}_{\rm{b}}, 
\end{align}

\noindent where $\vec{X}_{\rm{a}}$ need to be equal to $\vec{X}_{\rm{b}}$.  We needed to find the $\lambda_{\rm{a}}$ and $\lambda_{\rm{b}}$ values that satisfy the following condition:

\begin{align}
    \lambda_{\rm{a}} &= \frac{n^{\rm{y}}_{\rm{a}}\left(r^{\rm{x}}_{\rm{a}}-r^{\rm{x}}_{\rm{b}} \right)-n^{\rm{x}}_{\rm{b}}\left(r^{\rm{y}}_{\rm{s}}-r^{\rm{y}}_{\rm{b}} \right)}{n^{\rm{y}}_{\rm{a}}n^{\rm{x}}_{\rm{b}}-n^{\rm{x}}_{\rm{a}}n^{\rm{y}}_{\rm{b}}}\\
    \lambda_{\rm{b}} &= \frac{n^{\rm{y}}_{\rm{b}}\left(r^{\rm{x}}_{\rm{a}}-r^{\rm{x}}_{\rm{b}} \right)-n^{\rm{x}}_{\rm{a}}\left(r^{\rm{y}}_{\rm{a}}-r^{\rm{y}}_{\rm{b}} \right)}{n^{\rm{y}}_{\rm{a}}n^{\rm{x}}_{\rm{b}}-n^{\rm{x}}_{\rm{a}}n^{\rm{y}}_{\rm{b}}}.
\end{align}
\label{A8 and A9}

To calculate the angle between the RSs-RBs, we used the following equation:
\label{angle_eq}

\begin{equation}
    \theta_{\rm{a,b}} = \arccos\frac{\left(v^{\rm{a}}_{\rm{x}}v^{\rm{b}}_{\rm{x}} + v_{\rm{y}}^{\rm{a}}v_{\rm{y}}^{\rm{b}}\right)}{v^{\rm{a}}v^{\rm{b}}}.
    \label{eqn:Angle_Equation}
\end{equation}

By propagating the errors, we obtained a value corresponding to 1$\sigma$ of $\theta_{1\sigma}$. Then the upper and lower angles were obtained using

\begin{equation}
    \theta^+_{\rm{a,b}} = \arccos\frac{\left(v^{\rm{a}}_{\rm{x}}v^{\rm{b}}_{\rm{x}} + v_{\rm{y}}^{\rm{a}}v_{\rm{y}}^{\rm{b}}\right)}{v^{\rm{a}}v^{\rm{b}}} + \theta_{1\sigma}
    \label{eqn:Angle_Equation_Upper}
\end{equation}

\begin{equation}
    \theta^-_{\rm{a,b}} = \arccos\frac{\left(v^{\rm{a}}_{\rm{x}}v^{\rm{b}}_{\rm{x}} + v_{\rm{y}}^{\rm{a}}v_{\rm{y}}^{\rm{b}}\right)}{v^{\rm{a}}v^{\rm{b}}} - \theta_{1\sigma}
    \label{eqn:Angle_Equation_Lower}.
\end{equation}

The numerator of the fraction in the above equations, $(v^{\rm{a}}_{\rm{x}}v^{\rm{b}}_{\rm{x}} + v_{\rm{y}}^{\rm{a}}v_{\rm{y}}^{\rm{b}})$, represents the dot product of the velocity vectors of objects $a$ and $b$. The denominator, $v^{\rm{a}} v^{\rm{b}}$, represents the product of the magnitudes of the velocity vectors of objects $a$ and $b$. Taking the fraction $\frac{{v^{\rm{a}}_{\rm{x}}v^{\rm{b}}_{\rm{x}} + v_{\rm{y}}^{\rm{a}}v_{\rm{y}}^{\rm{b}}}}{{v^{\rm{a}} v^{\rm{b}}}}$, we obtain the cosine of the angle between the two velocity vectors. 

\end{appendix}


\label{LastPage}
\end{document}